\documentclass[preprint,3p,times]{elsarticle}

\usepackage[mathlines]{lineno}

\usepackage{amsmath, amssymb}
\usepackage{color}

\usepackage{hyperref}       
\usepackage{url}            
\usepackage{amsfonts}       
\usepackage{nicefrac}       
\usepackage{microtype}      
\usepackage{doi}
\usepackage{dcolumn}
\usepackage{bm}
\usepackage{caption}
\usepackage{subcaption}
\usepackage{float}
\usepackage{array}
\newcolumntype{P}[1]{>{\centering\arraybackslash}p{#1}}

\begin{document}

\begin{frontmatter}

\title{A Hybrid Deep Neural Operator/Finite Element Method for Ice-Sheet Modeling}

\author[UMN]{QiZhi He}
\author[Sandia]{Mauro Perego}
\author[PNNL]{Amanda A. Howard}
\author[PNNL,Brown]{George Em Karniadakis}
\author[PNNL]{Panos Stinis}

\address[UMN]{Department of Civil, Environmental, and Geo- Engineering, University of Minnesota, 500 Pillsbury Drive S.E., Minneapolis, MN 55455}

\address[Sandia]{Center for Computing Research, Sandia National Laboratories, P.O. Box 5800, Albuquerque, NM 87185}

\address[PNNL]{Advanced Computing, Mathematics and Data Division, Pacific Northwest National Laboratory, Richland, WA 99352}

\address[Brown]{Division of Applied Mathematics and School of Engineering, Brown University, 182 George Street, Providence, RI 02912}
\begin{abstract}
One of the most challenging and consequential problems in climate modeling is to provide probabilistic projections of sea level rise. A large part of the uncertainty of sea level projections is due to uncertainty in ice sheet dynamics. At the moment, accurate quantification of the uncertainty is hindered by the cost of ice sheet computational models. In this work, we develop a hybrid approach to approximate existing ice sheet computational models at a fraction of their cost. Our approach consists of replacing the finite element model for the momentum equations for the ice velocity, the most expensive part of an ice sheet model, with a Deep Operator Network, while retaining a classic finite element discretization for the evolution of the ice thickness. We show that the resulting hybrid model is very accurate and it is an order of magnitude faster than the traditional finite element model. Further, a distinctive feature of the proposed model compared to other neural network approaches, is that  it can handle high-dimensional parameter spaces (parameter fields) such as the basal friction at the bed of the glacier, and can therefore be used for generating samples for uncertainty quantification. We study the impact of hyper-parameters, number of unknowns and correlation length of the parameter distribution on the training and accuracy of the Deep Operator Network on a synthetic ice sheet model. We then target the evolution of the Humboldt glacier in Greenland and show that our hybrid model can provide accurate statistics of the glacier mass loss and can be effectively used to accelerate the quantification of uncertainty.
\end{abstract}




\end{frontmatter}

\section{Introduction}
Ice sheet models are important components of climate models and are crucial for providing projections of sea-level rise. In fact, sea-level rise is due in large part to added water to the ocean originating from mass loss of Greenland and Antarctic ice sheets \cite{IPCCSeaLevelChurch2013,levermann2020projecting,edwardsNature2021}. 

Quantifying the uncertainty on the projections of sea-level rise, due to uncertainties in the data and in the models, is an extremely challenging task. The large dimensionality of the parameter space, and high computational cost of ice sheet models make Bayesian inference and uncertainty quantification infeasible, despite the large computational resources available. While there are efficient ways to perform Bayesian inference under certain approximations \cite{isaac2015,brinkerhoff2022}, previous attempts to quantify the uncertainty on sea level rise (e.g., \cite{aschwanden2019, bulthuis2019, lehner2020}) perform drastic reductions of the dimensionality of the parameter space that are often dictated by feasibility reasons rather than by physical or mathematical arguments. 

Several efforts \cite{Bueler2009,Goldberg2011a,perego2012,Leng2012,Larour2012,Cornford2013,gagliardini2013,brinkerhoff2013,tezaur2015a, hoffman2018} over the last decades focused on efficiently solving the steady state Stokes-like flow equations governing the ice flow, which still represents the most computationally expensive part of an ice flow model. Flow equations need to be solved at each time step. While time steps can be as little as a week, typical temporal periods of interest range from a few decades to centuries, to millennia. 
In this work we aim at replacing the most expensive part of an ice sheet model, the Stokes-like flow equations, with a deep learning surrogate that is order of magnitudes faster than the  finite element based implementation. A similar idea has been pursued by Jouvet et al. \cite{jouvet2021}, where a deep learning model was used to accelerate ice sheet modeling of Paleo simulations. A key requirement for our surrogate, that sets it apart from \cite{jouvet2021}, is that it depends on high-dimensional parameter spaces (parameter fields), such as the basal friction coefficient that determines the basal sliding or the bed topography. This allows us to use the model for inference and for uncertainty quantification. We also note that in paleo simulations most of the uncertainty comes from the climate forcing, whereas in the simulations in which we are interested here, that spans approximately half a century, model error is a significant source of uncertainty \cite{lehner2020}, which forces us to have very accurate models. Another related problem, where deep learning models have been used to approximate the parameter-to-velocity map in ice-sheet problems, is presented in  \cite{brinkerhoff2021}. In that work, the authors first find a basis of the operator using principal component analysis, and then use a residual neural network to compute the basis coefficients as a function of the parameters. In contrast to our problem, in \cite{brinkerhoff2021} only a handful of parameters are considered.

We represent our deep learning surrogate with Deep Operator Networks (DeepONets) \cite{luluDeepONet2021}, which have proven to work well in learning operators in a wide range of applications ranging from fracture mechanics to combustion problems \cite{lin2021seamless, ranade2021generalized, goswami2021physics, di2021deeponet, sharma2021application}. In its vanilla formulation, a DeepONet contains two deep neural networks, referred to as the \emph{branch} network and the \emph{trunk} network. The trunk network takes as input spatial coordinates whereas the branch network takes as input the input fields evaluated at a fixed set of points. DeepONets approximate operators as a linear combination of ``basis functions'' generated by the trunk network, with coefficients generated by the branch network. The mathematical foundations of DeepONets are based on the universal approximation theorem \cite{chen1995universal, back2002universal}, and, under mild assumptions, it has been proven that DeepONets can approximate with given accuracy any operator \cite{luluDeepONet2021}.  Our DeepONet surrogate takes as input fields the ice thickness and the basal friction field and computes the depth-averaged ice velocity field. 

We use the trained DeepONet to build a fast \emph{hybrid} ice-flow model, where the evolution of the ice thickness is discretized with a classic finite element method, and, at each time step, the ice velocity field (as a function of the ice thickness and the basal friction field) is computed by the DeepONet. A finite element implementation of the ice-flow model is used as the ``reference model'' and also used to generate data to train the the DeepONet model. 
We demonstrate our approach on two ice sheet problems: 1. a synthetic ice sheet problem for exploring different hyper-parameters of the DeepONet and for studying the impact of mesh resolution and correlation length on the DeeoONet training and accuracy, and 2. a realistic simulation of the Humboldt glacier, which is one of the largest glaciers in Greenland and one that is expected to greatly contribute to sea-level rise in this century \cite{Hillebrand2022}. We show how our DeepONet surrogate can approximate the ice velocity computed by the finite element model very accurately (relative error of 0.4\%) and at a fraction of the cost of the finite element model. The hybrid model produces accurate results for the ice thickness (\~ 2\% relative error over a span of 100 years). We also show how the mass loss of the Humboldt glacier, computed using the hybrid model, is an accurate representation of the finite element model and can be used for computing statistics of sea level rise, yielding a 10 fold speed-up.

In Section \ref{sc:models} we present the mathematical equations that we use to compute the ice thickness and velocity and the probability distribution of the basal friction parameter.  In section \ref{sc:computational_models} we introduce the hybrid model, focusing in particular on its DeepONet component.  In Section \ref{sc:MISMIP} we present the result of training the DeepOpNet for a synthetic test case, studying how the resolution of the input data and the correlation length of the basal friction distribution affect the accuracy and training of the DeepONet. Finally in Section \ref{sc:Humboldt} we target the Humboldt glacier and show how hybrid model can be effectively used for computing the statistics of the glacier mass loss. We conclude in Section \ref{sc:summary} with a summary.

\section{Ice Sheet Models} \label{sc:models}
\begin{figure}[h]
   \begin{center}
   \includegraphics[width=0.7\textwidth]{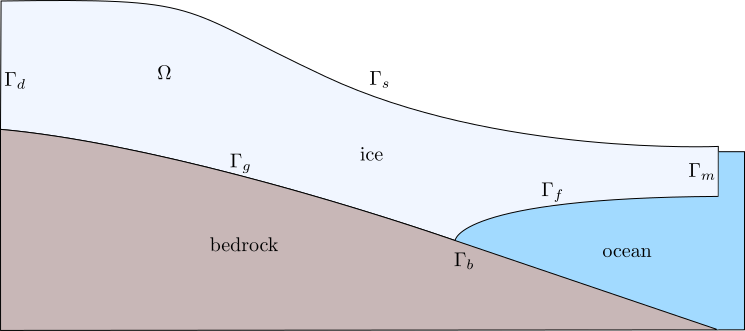}
   \end{center}
   \caption{Cartoon of an ice sheet in the $x-z$ plane.} \label{fig:iceshelf}
\end{figure}

In this section, we briefly introduce the ice sheet models considered in this work, as depicted in Fig. \ref{fig:iceshelf}.

Let $x$ and $y$ denote the horizontal coordinates and $z$ the vertical coordinate, chosen such that the sea level corresponds to $z=0$.
 The ice domain, at time $t$, can be approximated as a vertically extruded domain $\Omega$ defined as 
$$
\Omega(t) := \{(x,y,z) \; \text{s.t.} \; (x,y) \in \Sigma, \; \text{and}\; l(x,y,t) < z < s(x,y,t) \},
$$
where $\Sigma \subset \mathbb R^2$ is the horizontal extension of the ice. 
$\Gamma_l(t)  := \{(x,y,z) \; \text{s.t.} \; z = l(x,y,t)\}
$ denotes the lower surface of the ice at time $t$, and $\Gamma_s(t)  := \{(x,y,z) \; \text{s.t.} \; z = s(x,y,t)\}$ denotes the upper surface of the ice\footnote{For simplicity here we assume that $\Sigma$ does not change in time. This implies that the ice sheet cannot extend beyond $\Sigma$ but it can become thicker or thinner (to the point of disappearing in some regions).}. The bed topography, which we assume constant in time, is given by $\Gamma_b := \{(x,y,z) \; \text {s.t.} \; z = b(x,y)\}$. In general, the ice sheet can have ice shelves where the ice is floating. We hence partition the lower surface of the ice $\Gamma_l$ in the grounded part $\Gamma_g = \Gamma_l \cap \Gamma_b$ (here, $l(x,y,t) = b(x,y)$) and the floating part $\Gamma_f$ under the ice shelf. We partition the lateral boundary of $\Omega$ in $\Gamma_m$, denoting the ice sheet margin (either terrestrial or marine margin), and, when we only consider a portion of the ice sheet, in $\Gamma_d$, denoting an internal (artificial) boundary often chosen in correspondence of the ice divides.

The thickness of the ice, given by $H(x,y,t) := s(x,y,t)-l(x,y,t)$,  is defined on $\Sigma \times [0, t_f]$ and evolves according to
\begin{equation} \label{thickness}
  \partial_t H + \nabla \cdot (\mathbf{\bar u} H) = f_H 
\end{equation}
where $\displaystyle \mathbf{\bar u} := \frac{1}{H} \int_l^s \mathbf{u}\, dz$ is the depth-integrated velocity and $f_H$ is an accumulation rate, accounting for accumulation (e.g., due to snow precipitations) and melting at the upper surface and accumulation/melting at the base of the ice sheet. 
We need to constrain $H$ to be non-negative, as there is no guarantee that $f_H$, typically coming from climate models, is consistent with the ice thickness equation. 

Ice sheets behave as a shear thinning fluid and can be modeled with the nonlinear Stokes equation \cite{Cuffey2010}. In this work we use simplifications of Stokes equations that are less expensive to solve and that are obtained with scaling arguments based on the fact that glaciers and in particular ice sheets are typically shallow. We consider two such simplifications: the mono-layer higher-order approximation (MOLHO)  and the shallow shelf approximation (SSA). The MOLHO model \cite{dosSantos2022} is suitable for both frozen and thawed beds, whereas the simpler SSA model \cite{morland_johnson_1980, weis_1999} works well only for grounded ice with significant sliding at the bed or for ice shelves where the ice is floating over the water. 
In the following we detail the Stokes model and its approximations.
\subsection{Stokes model}
We denote with $u$, $v$ and $w$ the $x$, $y$ and $z$ components of the ice velocity, respectively, and the ice velocity vector is denoted by $\boldsymbol{u} := (u,v,w)$. Denoting the pressure with $p$, and the ice density with $\rho$, the Stokes equation reads  
\begin{align}
    -\nabla \cdot \sigma & = \rho \mathbf{g} \\
    \nabla \cdot \mathbf{u} &= 0
\end{align}
with stress tensor $\sigma = 2\mu \mathbf{D} - pI$, and strain rate tensor  $\mathbf{D}_{ij}(\mathbf{u}) = \frac{1}{2} \left(\frac{\partial u_i}{\partial x_j}+\frac{\partial u_j}{\partial x_i} \right)$.
The non-linear viscosity is given by 
\begin{equation} \label{viscosity}
    \mu = \frac12 A(T)^{-q} \, D_e(\mathbf u)^{q-1}
\end{equation}
with $q\leq 1$. In this work we take $q = \frac{1}{3}$, a typical choice. $A$ is the ice flow factor that depends on the ice temperature $T$. The effective strain rate $D_e(\mathbf u)$ 
is given by $D_e(\mathbf u) = \frac{1}{\sqrt{2}}|\mathbf{D}(\mathbf{u})|$, where $| \cdot |$ denotes the Frobenius norm.
The Stokes equation is accompanied by the following boundary conditions:
$$
\left \{ \begin{array}{lll}
\sigma \mathbf{n} = 0 & \text{on } \Gamma_s & \text{stress free, atmospheric pressure neglected} \\
\sigma \mathbf{n} = \rho_w \, g\,  \min(z,0) \bf{n} & \text{on } \Gamma_{m}  & \text{boundary condition at the ice margin} \\
\mathbf u = \mathbf u_d & \text{on } \Gamma_{d}  & \text{Dirichlet condition at internal boundary} \\
\bf{u} \cdot \bf{n} = 0,\; (\sigma \bf{n})_{\parallel} = \beta \bf{u}_{\parallel} & \text{on } \Gamma_g & \text{impenetrability + sliding condition} \\
\sigma \mathbf{n} = \rho_w \, g\,  z\, \bf{n} & \text{on } \Gamma_f & \text{back pressure from ocean under ice shelves} \\
\end{array} \right.
$$
Here $\beta(x,y)$ is the sliding (or friction) coefficient, $\rho_w$ is the density of the ocean water and $\mathbf{n}$ the unit outward-pointing normal to the boundary. The boundary condition at the margin includes the ocean back-pressure term, when the margin is partially submerged ($z<0$). For terrestrial margin, $z>0$, hence the term becomes a stress-free condition. The friction term $\beta$ can also depend on $\mathbf{u}$, depending on the choice of the sliding law.  

\subsection{Mono-layer higher-order (MOLHO)} \label{sec:molho}
The MOLHO model~\cite{dosSantos2022} is based on the Blatter-Pattyn approximation~\cite{Dukowicz2010} that can be derived neglecting the terms $w_x$ and $w_y$ in the strain-rate tensor $D$ and, using the continuity equation, replacing $w_z$ with $-(u_x+v_y)$:
\begin{equation}
    \mathbf{D} = \begin{bmatrix} u_x  & \frac12(u_y + v_x) & \frac12 u_z \\[2mm]
    \frac12(u_y + v_x) & v_y & \frac12 u_z \\[2mm]
        \frac12 u_z & \frac12 v_z & -(u_x+v_y) \end{bmatrix}.
\end{equation}
This leads to the following elliptic equations in the horizontal velocity $(u,v)$
\begin{equation}
    -\nabla \cdot (2\mu \hat{\mathbf{D}} ) = - \rho g \nabla s
\end{equation}
with 
\begin{equation}
    \hat{\mathbf{D}} = \begin{bmatrix} 2u_x + v_y & \frac12 (u_y + v_x) & \frac12 u_z \\
    \frac12 (u_y + v_x) & u_x + 2v_y & \frac12 v_z \end{bmatrix}.
\end{equation}
Here the gradient is two-dimensional: $\nabla = [\partial_x, \partial_y]^T$. The viscosity $\mu$ is given by \eqref{viscosity} with the effective strain rate 
$$
D_e = \sqrt{u_x^2 + v_y^2 + u_x v_y + \frac{1}{4} (u_y + v_x)^2 + \frac{1}{4} u_z^2 + \frac{1}{4} v_z^2}.
$$
The boundary conditions reads
$$
\left \{ \begin{array}{lll}
2\mu \hat{\mathbf{D}}\, \mathbf{n} = 0 & \text{on } \Gamma_s & \text{stress free, atmospheric pressure neglected} \\
2\mu \hat{\mathbf{D}}\, \mathbf{n} = \psi \bf{n} & \text{on } \Gamma_{m}  & \text{boundary condition at at ice margin} \\
\mathbf u = \mathbf u_d & \text{on } \Gamma_{d}  & \text{Dirichlet condition at internal boundary} \\
2\mu \hat{\mathbf{D}}\, \mathbf{n} = \beta \bf{u}_{\parallel} & \text{on } \Gamma_g & \text{sliding condition} \\
2\mu \hat{\mathbf{D}}\, \mathbf{n} = 0 & \text{on } \Gamma_f & \text{free slip under ice shelves} \\
\end{array} \right.
$$
where $\psi = \rho g (s-z) \mathbf{n} + \rho_w \, g\,  \min(z,0) \bf{n}$, which can be approximated with its depth-averaged value  $\bar{\psi} = \frac12 g H (\rho - r^2 \rho_w)$, $r$ being the the submerged ratio $r=\max\left(1-\frac{s}{H},0\right)$; $ \bf{u}_{\parallel}$ is the component of the velocity $ \bf{u}$ tangential to the bed.

MOLHO consists of solving the weak form of the Blatter-Pattyn model, with the ansatz that the velocity can be expressed as :
$$
\mathbf u(x,y,z) = \mathbf u_b(x,y) + \mathbf u_v(x,y) \left( 1 - \left(\frac{s-z}{H}\right)^{\frac{1}{q}+1}\right).
$$
The problem is then formulated as a system of two two-dimensional partial differential equations (PDEs) for $\mathbf u_b$ and $\mathbf u_v$ (for a detailed derivation see~\cite{dosSantos2022}.) Note that the depth-averaged velocity is given by
$\mathbf{\bar u} = \mathbf u_b + \frac{(1+q)}{(1+2q)} \;\mathbf u_v$.

\subsection{Shallow Shelf Approximation (SSA)} \label{sec:ssa}
The shallow shelf approximation~\cite{morland_johnson_1980} is a simplification of the Blatter-Pattyn model, assuming that the velocity is uniform in $z$, so $\mathbf{u}=\mathbf{\bar u}$. It follows that $u_z=0$ and $v_z=0$, giving:

\begin{equation}
    \mathbf{D} = \begin{bmatrix} u_x  & \frac12 (u_y + v_x) & 0 \\
    \frac12 (u_y + v_x) & v_y & 0 \\ 
        0 & 0 & -(u_x+v_y) \end{bmatrix}, \quad 
    \hat{\mathbf{D}} = \begin{bmatrix} 2u_x + v_y & \frac12 (u_y + v_x) & 0 \\
    \frac12 (u_y + v_x) & u_x + 2v_y & 0 \end{bmatrix}, 
\end{equation}
and $D_e = \sqrt{u_x^2 + v_y^2 + u_x v_y + \frac{1}{4} (u_y + v_x)^2}$.
The problem simplifies to a two-dimensional PDE in $\Sigma$
$$
- \nabla \cdot \left( 2 \mu H  \hat {\mathbf D }(\mathbf{\bar u})\right) + \beta \mathbf{\bar u} =  -  \rho g H  \nabla s, \quad \text{in } \Sigma
$$
with $\bar \mu = \frac12 \bar{A}(T)^{-\frac{1}{n}} \, D_e(\mathbf{\bar u})^{\frac{1}{n}-1}$, where $\bar{A}$ is the depth-averaged flow factor and with boundary conditions:
$$
\left \{ \begin{array}{lll}
2\mu \hat{\mathbf{D}}(\mathbf{\bar u})\, \mathbf{n} = \bar \psi \bf{n} & \text{on } \Gamma_{m}  & \text{boundary condition at ice margin} \\
\mathbf{\bar u} = \mathbf{\bar u}_d & \text{on } \Gamma_{d}  & \text{Dirichlet condition at internal boundary} \\
\end{array} \right .
$$
Recall that $\bar{\psi} = \frac12 g H (\rho - r^2 \rho_w)$, $r$ being the the submerged ratio $r=\max\left(1-\frac{s}{H},0\right)$. With abuse of notation, here $\Gamma_{m}$ and $\Gamma_d$ are intended to be subsets of $\partial \Sigma$. 

\subsection{Distribution of basal friction field}\label{sec:friction}
The basal friction field $\beta$ is one of the main factors that control the ice velocity. It cannot be measured directly and it is typically estimated by solving a PDE-constrained optimization problem, e.g., \cite{perego2014, perego2022}, to assimilate observations of the surface ice velocity. As a result, the basal friction field is affected by both uncertainties in the observations and in the the model. While it is possible to characterize the probability distribution for $\beta$ using a Bayesian inference approach, e.g., \cite{petra2014}, here we adopt a simplified log-normal distribution for $\beta$. We write the basal friction field as $\beta = \exp(\gamma)$, where $\gamma$ is normally distributed as
\begin{equation} \label{distribution}
\gamma \sim \mathcal F\left(\log(\bar{\beta}), k_l\right),\; \text{ and } \; k_l(\mathbf x_1,\mathbf x_2) = a \exp\left(-\frac{|\mathbf x_1-\mathbf x_2|^2}{2l^2}\right).
\end{equation}
 Here $\log (\bar{\beta})$ is the mean of the Gaussian process $\mathcal F$ and it is often obtained by assimilating the observed velocities \cite{perego2014}, $l$ is the correlation length and $a$ is a scaling factor. In this work we choose values of the correlation length and of the scaling factor that produce reasonable results. While an in-depth validation of the chosen parameters is beyond the scope of this work, we explore the dependence of the accuracy of the DeepONet model as a function of the correlation length, as discussed in Section \ref{sc:MISMIP}.
\section{Computational Models} \label{sc:computational_models}
In this section we introduce the finite element ice flow model and the hybrid ice flow model. We first perform a semi-implicit time discretization of the ice thickness equation \eqref{thickness}:
\begin{equation} \label{time-disc-thickness}
\left \{ \begin{array}{lll}
H^{n+1} &=& H^{n} - \Delta t \, \nabla \cdot \left ( \mathbf{\bar u}^n H^{n+1}\right) + \Delta t F_H^n \\
\mathbf{\bar u}^n &=& \mathcal G(\beta, H^n)
\end{array} \right.
\end{equation} 
where $H^n$ is the approximation of $H$ at time $t^n = t^0 + n \Delta t$, for a given time-step $\Delta t$, and $F_H^n = F_H(t^n)$ is the corresponding discrete approximation of the accumulation rate $f_H$. Here, $\mathcal G(\cdot,\, \cdot)$ is the velocity operator that maps the basal friction field and the ice thickness into the depth-averaged velocity vector, based either on the SSA (Sec. \ref{sec:ssa}) model or the MOLHO (Sec. \ref{sec:molho}) model. 
In this work we discretize the thickness equation \eqref{time-disc-thickness} with finite elements, using streamline upwind stabilization. Similarly, we provide a classic Galerkin finite element discretization of the nonlinear operator  $\mathcal G$. The finite element discretization is implemented in \texttt{FEniCS}~\cite{fenics2015}. We use continuous piece-wise linear finite elements for both the thickness and the velocity fields, and solve the discretized problem with \texttt{PETSc}~\cite{petsc1998} SNES nonlinear solvers. We refer to this finite element implementation of \eqref{time-disc-thickness} as the \emph{finite element ice flow model} that we use as our a reference model. 

The focus of the paper is on avoiding the high computational cost of constructing a finite element approximation of the nonlinear operator $\mathcal G$, and using, instead, a DeepONet approximation of $\mathcal G$, which, in combination with the finite element discretization of the first equation of \eqref{time-disc-thickness}, constitutes the \emph{hybrid ice flow model}. The DeepONet implementation and training are performed using \texttt{JAX}~\cite{jax2018}. At each time step, the \texttt{FEniCS} finite element code calls the\texttt{JAX} DeepONet code to compute an approximation of $\mathcal G(\beta, H^n)$. In the next  sections we describe in detail the DeepONet architecture and its training.

\subsection{DeepONet approximation}
\begin{figure}[ht]
\centering
\includegraphics[width=0.8\textwidth]{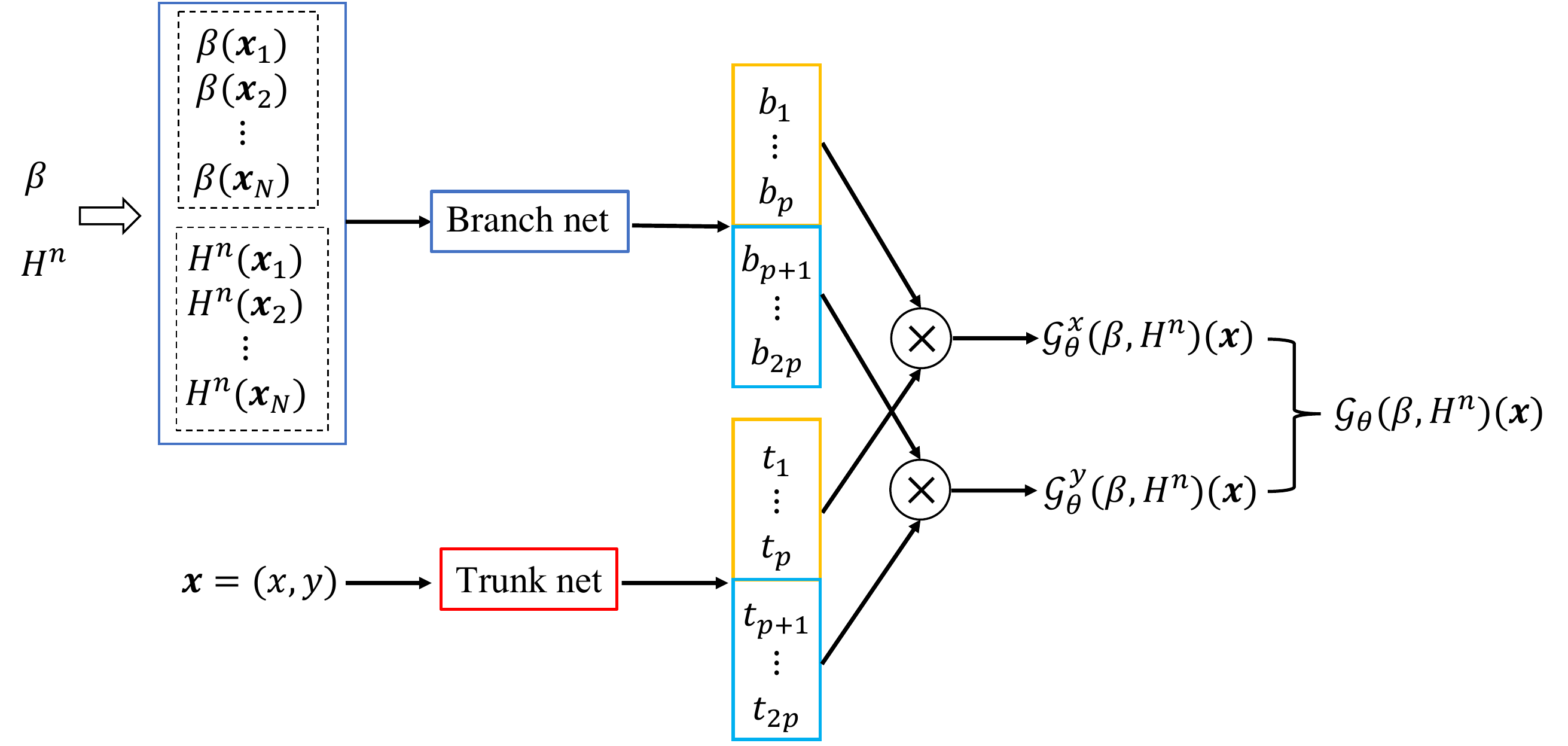}
\caption{Schematic representation of DeepONet. The branch net takes as inputs the functions $\beta (\mathbf{x})$ and $H^n(\mathbf{x})=H(t_n,\mathbf{x})$ evaluated at $N$ fixed sensor points $ \mathcal{X} = \{\mathbf{x}_i\}_{i=1}^N$ and returns the feature embedding vector $ \mathbf{b} \in \mathbb{R}^{2p}$ as output. The trunk net takes the continuous coordinates $\mathbf{x} \in \mathcal{Y}$ as input and outputs another embedding vector $ \mathbf{t} \in \mathbb{R}^{2p}$. The embedding vectors $ \mathbf{b}$ and $ \mathbf{t}$ are combined by dot product to generate the solution operator, $\mathcal{G}_{\theta} (\beta, H^n)(\mathbf{x})$. The trainable parameters $\theta$ associated with the branch net and the trunk net are optimized by minimizing the loss function defined as a weighted mean square error (see Eq. \ref{eq:onet_loss}). In this study, we set $\mathcal{Y} = \mathcal{X}$ for simplicity.}
\label{fig:deeponet}
\end{figure}

As briefly discussed in the introduction, the main idea of DeepONet is to learn, in general nonlinear, operators mapping between infinite-dimensional function spaces via deep neural networks \cite{luluDeepONet2021}. Inspired by the universal approximation theorem for operators~\cite{chen1995universal}, DeepONet's architecture consists of two neural networks: one is used to encode the input function sampled at fixed sensor points (\textit{branch net}) whereas the other inputs the location coordinates to evaluate the output function (\textit{trunk net}). It has been shown that this architecture of two sub-networks can substantially improve generalization compared to fully connected neural networks \cite{luluDeepONet2021}. In this study, a DeepONet denoted by $ \mathcal{G}_\theta$ is used as a surrogate  for the nonlinear operator $\mathcal G$ in Eq. \eqref{time-disc-thickness},
\begin{equation}
    \mathcal{G}_\theta (\beta,H^n)(\mathbf{x}) \approx  \mathcal{G} (\beta,H^n)(\mathbf{x}),
\end{equation}
where $\theta$ represents the collection of trainable parameters in DeepONet, and the approximated velocity components are
\begin{align}
\begin{split}
& \bar{u}^{n}_x \approx \mathcal{G}^x_\theta(\beta,H^n)(\mathbf{x}) = \sum_{m=1}^{p} b_m (\beta,H^n) t_m (\mathbf{x}), \\
& \bar{u}_y^{n} \approx \mathcal{G}^y_\theta (\beta,H^n)(\mathbf{x}) = \sum_{m=p+1}^{2p} b_m (\beta,H^n) t_m (\mathbf{x}),
\end{split}
\end{align}
where $b_m$ and $t_m$ denote the outputs of the branch net and the trunk net, respectively. 
The details of the DeepONet model is shown in the schematic of Fig. \ref{fig:deeponet}.
In this setting, the input functions, i.e., the friction $\beta$ and thickness $H^n$ at the moment $t^n$, evaluated at finite locations (sensors), $\mathcal{X} = \{\mathbf{x}_1,\mathbf{x}_2,...,\mathbf{x}_N\}$, are mapped as embedded coefficients through the branch net, while the trunk net learns a collection of space-dependent basis functions that are linearly combined with the branch coefficients to  approximate the velocity components. Note that the learned operator  $\mathcal{G}_\theta (\beta,H^n)$ is a continuous function with respect to coordinates $\mathbf{x}$, which are the inputs to the trunk net.
For brevity, we denote the DeepONet approximated velocity as $\bar{\mathbf{u}}^{NN}$.

\subsection{DeepONet training} \label{sc:training}
The trainable parameters, i.e., $\bm{\theta}$, associated with the DeepONet model are obtained by minimizing the loss function
\begin{equation}\label{eq:onet_loss}
\mathcal{L} (\bm{\theta}) = \frac{1}{N_{\beta} N_T} \sum_{i=1}^{N_{\beta}} \sum_{j=1}^{N_{T}} \sum_{\mathbf{x} \in \mathcal Y} w_{ij} (\mathbf{x}) | \mathbf{\bar{u}}(\mathbf{x},t^j;\beta_i) -  \mathcal{G}_\theta (\beta_i,H^j)(\mathbf{x})|^2,
\end{equation}
where $w_{ij} (\mathbf{x})$ are weights corresponding to each data point, $N_{\beta}$ is the number of friction fields $\{\beta_i\}_{i=1}^{N_\beta}$ used for different training simulations, $N_{T}$ is the number of time steps within each simulation to sample the velocity and thickness, $\mathbf{\bar{u}}(\mathbf{x},t^j;\beta_i)$ is the target velocity solution, and $\mathcal{G}_\theta (\beta_i,H^j)(\mathbf{x})$ is the predicted value obtained from DeepONet. Both target solution $\mathbf{\bar{u}}(\mathbf{x},t^j;\beta_i):= \mathcal G(\beta_i, H^j) (\mathbf{x})$ and DeepONet prediction $\mathcal{G}_\theta (\beta_i,H^j)(\mathbf{x})$ are evaluated at the set of locations $\mathcal{Y}$. The input functions $\beta_i$ and $H^j$ of the branch network are discretized at the fixed set of sensor points $\mathcal{X}$ (see Fig. \ref{fig:deeponet}). In this work it is convenient to choose $\mathcal{X}$ to be the set of the grid nodes used in the finite element discretization and to take $\mathcal{Y} = \mathcal{X}$.

In Eq. \eqref{eq:onet_loss}, the penalizing weights $w_{ij} (\mathbf{x})$ are generally related to the characteristics of training data, i.e., the friction field, time step, and spatial locations. For simplified cases where the target operator presents little variability with respect to the input parameters, the weights are assumed to be unity, i.e., $w_{ij} (\mathbf{x}) \equiv 1$. However, it is observed in our numerical investigation that using nonuniform (space-dependent) weights can lead to better generalization. 
To this end, we use the self-adaptive weight estimation approach~\cite{mcclenny2020self,goswami2022physics} to adjust the weight parameters through gradient descent along with the network parameters. Assuming that the weights depend only on the space coordinates, i.e., $w_{ij} (\mathbf{x}) = w (\mathbf{x})$, the loss function \eqref{eq:onet_loss} is modified as
\begin{equation}\label{eq:onet_loss_SA}
\mathcal{L} (\bm{\theta}, \bm{\lambda})= \frac{1}{N_{\beta} N_T} \sum_{i=1}^{N_{\beta}} \sum_{j=1}^{N_{T}} \sum_{\mathbf{x} \in \mathcal Y} w (\mathbf{x}) | \mathbf{\bar{u}}(\mathbf{x},t^j;\beta_i) -  \mathcal{G}_\theta (\beta_i,H^j)(\mathbf{x})|^2,
\end{equation}
where $ w (\mathbf{x})$ is further defined as $ m(\lambda (\mathbf{x}))$ in which $\bm{\lambda} = \{\lambda(\mathbf{x})\}_{\mathbf{x} \in \mathcal Y}$ are the trainable self-adaptive weight parameters dependent on locations $\mathbf{x}$, and $m(\lambda)$ is a mask function defined on $[0, \infty]$ to accelerate convergence \cite{mcclenny2020self}.
The mask function needs to be differentiable, nonnegative, and monotonically increasing. The polynomial mask $m(\lambda) = \lambda^q$ for $q = 1,2,...$ is adopted in this study. 

The key feature of self-adaptive DeepONet training is that the loss $\mathcal{L} (\bm{\theta}, \bm{\lambda})$ is simultaneously minimized with respect to the network parameters $\bm{\theta}$ but maximized with respect to the self-adaptive parameters $\bm{\lambda}$, i.e.,
\begin{equation}\label{eq:onet_loss_SA_minmax}
\min_{\bm{\theta}} \max_{\bm{\lambda}} \mathcal{L} (\bm{\theta}, \bm{\lambda}).
\end{equation}
If one uses the gradient descent method, the updated equations of the two sets of parameters at $v$ iteration are:
\begin{align}\label{eq:SA_update}
\begin{split}
\bm{\theta}^{v+1} = \bm{\theta}^{v} - \eta_\theta \nabla_\theta \mathcal{L} (\bm{\theta}^{v}, \bm{\lambda}^{v}), \\
\bm{\lambda}^{v+1} = \bm{\lambda}^{v} + \eta_\lambda \nabla_\lambda \mathcal{L} (\bm{\theta}^{v}, \bm{\lambda}^{v}),
\end{split}
\end{align}
where $\eta_\theta$ and $\eta_\lambda$ are the learning rates for updating $\bm{\theta}$ and $\bm{\lambda}$, respectively. The employment of self-adaptive weights can significantly improve the prediction accuracy at the localized features in the solution by properly balancing the terms via the corresponding weights \cite{mcclenny2020self,kontolati2022influence}.

\subsection{Data preparation \& training details}
To generate sufficient training data, we perform simulations of the finite element ice flow model \eqref{time-disc-thickness} based on either SSA or MOLHO and considering $N_{\beta}$ basal friction samples, $\beta_i (\mathbf{x})$, $i=1,...,N_\beta$, taken from distribution \eqref{distribution}. For each sample $\beta_i$, we compute the thickness and depth-integrated velocity using the finite element flow model and  store their values $\{H_i^j\}_{j=1}^{N_T}$ and $ \{\mathbf{\bar{u}}_i^{j} \}_{j=1}^{N_T}$ at times  $t^j$, $j=1,2,...,N_T$ and grid points $\mathbf{x}_i \in \mathcal X$.

In training the DeepONet, the input functions, $\beta (\mathbf{x})$ and $H^n(\mathbf{x})$, as well as the DeepONet operator $\mathcal{G}_{\theta}$ are evaluated at points $\mathcal{X}=\{\mathbf{x}_1,\mathbf{x}_2,...,\mathbf{x}_N\}$, as described in Fig.~\ref{fig:deeponet}.
Therefore, a DeepONet training dataset is expressed as a triplet of the form,
\begin{equation}
\left[ \left \{ [\bm{\beta}^{(k)}, \bm{H}^{(k)} ] \right \}_{k=1}^{N_{\beta} N_{T}}, \left \{ \mathcal{Y}^{(k)} \right \}_{k=1}^{N_{\beta} N_{T}}, \left \{\mathbf{\bar{U}}^{(k)} \right \}_{k=1}^{N_{\beta} N_{T}} \right],
\end{equation}
where
\begin{align}
\begin{split}
& [\bm{\beta}^{(k)}, \bm{H}^{(k)} ] = [\beta_j(\mathbf{x}_1), \beta_j (\mathbf{x}_2),...,\beta_j(\mathbf{x}_N),
                                    H_i^j(\mathbf{x}_1), H_i^j(\mathbf{x}_2),..., H_i^j(\mathbf{x}_N)] , \\    
&  \mathcal{Y}^{(k)}  \equiv  \mathcal{X} = \{\mathbf{x}_1,\mathbf{x}_2,...,\mathbf{x}_{N}\}, \\
&  \mathbf{\bar{U}}^{(k)} = [\mathbf{\bar{u}}_i^{j} (\mathbf{y}_1),\mathbf{\bar{u}}_i^{j} (\mathbf{y}_2),...,\mathbf{\bar{u}}_i^{j} (\mathbf{y}_{N_{\mathbf u}})].
\end{split}
\end{align}
Here, the superscript $k$ is defined as $k = (i-1) N_T + j$ with $i = 1,...,N_{\beta}$ and $j = 1,...,N_T$, denoting the index of input parameters associated with time steps and friction samples. 

Regarding the basal friction fields, we adopt the following procedure to split the training and testing data: if $N_b$ friction fields are generated from the Gaussian process described in Section \ref{sec:friction}, the simulation solutions associated with the first $20$ fields, $\{ \beta_i \}_{i=1}^{20}$, are exclusively used for testing, while the rest $N_\beta =N_b -20$ fields, $\{ \beta_i \}_{i=21}^{20 + N_\beta}$, are selected for training the DeepONet model.
Unless stated otherwise, for the given training basal friction fields the finite element solutions at time steps $t = 1, 2, ..., 100$ (i.e., $N_T = 100$) are used for the training. 




In the following tests, the default training scheme uses the Adam optimizer with a learning rate $1\times 10^{-3}$. ReLU is selected as the activation function, and the batch size is $200$. 
The architecture of both the branch net and the trunk net is a fully connected neural network consisting of 4 hidden layers and 300 neurons per layer (denoted as $4 \times 300$).
To mitigate possible overfitting in training, we also introduce an $\ell^2$ regularization in \eqref{eq:onet_loss_SA} with a small penalty coefficient $5 \times 10^{-5}$.
However, we note that we did not observe any signs of conventional overfitting during our numerical tests, 
and the additional regularization has a negligible impact on the DeepONet accuracy.


\section{Synthetic Ice-Sheet Problem}\label{sc:MISMIP}
In this section we apply our approach to a well-known benchmark in ice sheet modeling, the MISMIP problem \cite{Cornford2020}. We use this problem to explore how hyper-parameters affect the training of the DeepONet and the accuracy of the hybrid model.

The problem geometry is defined by a marine ice stream that is partially floating. The ice domain is 640~km long and 80~km wide ($\Omega = [0,\,640\,\text{km}]\times[0,\,80\,\text{km}]$). The bed topography is provided in \cite{Cornford2020}. We consider an initial thickness (note that this is different from the one in \cite{Cornford2020}):
 $$
 H(x,y) = 100\,\text{m} \left(\frac{3}{2} + \frac{1}{2}\tanh{\left(\frac{400\,\text{km} - x}{100\,\text{km}}\right)}\right).
 $$
 We prescribe the normal velocity at the upstream boundary ($x=0\,$km) and lateral boundaries ($y=0\,$km and $y=80\,$km)  to be zero, and free-slip conditions in the direction tangential to these boundaries. We prescribe stress-free conditions at the outlet boundary ($x=640\,$km). No boundary conditions are prescribed for the thickness equation, as there are no inflow boundaries.
 We use a constant mean basal friction field $\bar \beta = 5000\,$Pa yr/m and a scaling factor $a = 0.2$ in \eqref{distribution}. As described in Section \ref{sc:training}, for each sample $\beta$ from \eqref{distribution}, we run the finite-element ice flow model for $100$ years, using a constant forcing $f_H = 0.3$ m/ yr, and compute the ice thickness $H$. We then use the thickness data to train the DeepONet.

For ease of analysis, the mean squared error (MSE) and relative squared error (RSE), given as follows, are used to evaluate the DeepONet performance:
 $$
    e_{MSE} = \frac{1}{N} \sum_{i=1}^{N} ||u_i - u_i^*||^2, \quad e_{RSE} = \frac{\sum_{i=1}^{N} ||u_i - u_i^*||^2}{\sum_{i=1}^{N} ||u_i^*||^2} 
 $$
 where $u_i$ and $u_i^*$ denote the prediction and reference values, respectively, and $N$ is the number of data.
\begin{figure}[ht]
\centering
\includegraphics[width=1\textwidth]{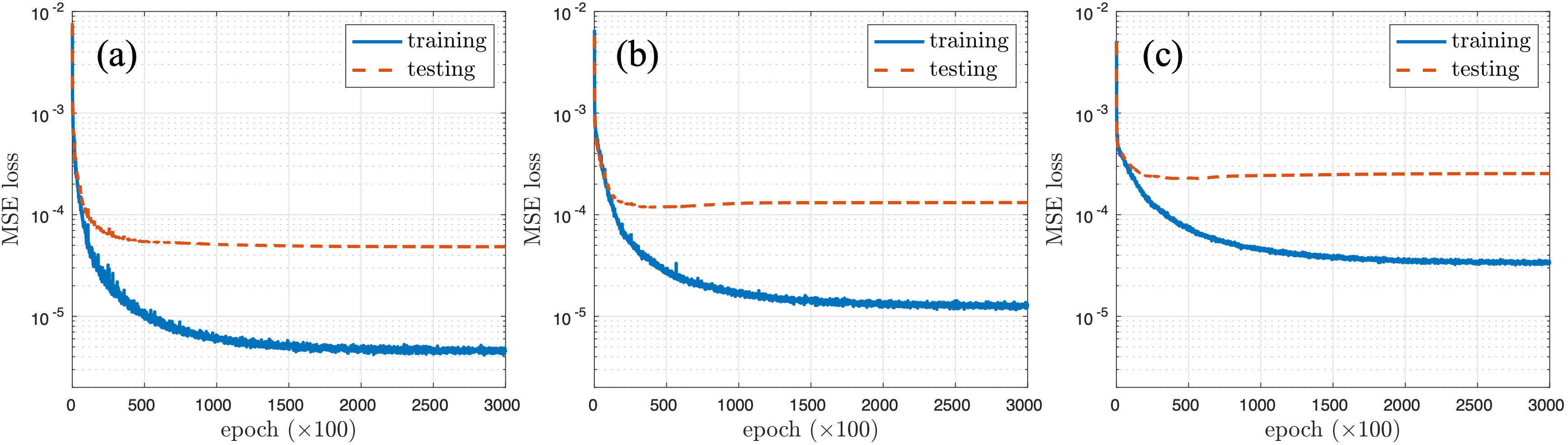}
\caption{The loss plots of DeepONet training for the MISMIP testcase with SSA model under different correlation lengths: 
(a) $l = 80$ km; (b) $l = 40$ km; (c) $l = 20$ km. The simulation data associated with $\{\beta_i\}_{i=21}^{300}$ is used as the training data while $\{\beta_i\}_{i=1}^{20}$ is used as testing data. At the final epoch ($300,000$), the training MSEs are $2.60 \times 10^{-6}$, $1.03 \times 10^{-5}$, and $3.33 \times 10^{-5}$, respectively.}
\label{fig:mismip-traning}
\end{figure}


\begin{table}[htbp]
\footnotesize
\caption{MISMIP test case with the SSA and MOLHO models. The mean square errors of the DeepONet training with different training dataset sizes under various correlation lengths $l$.
The testing error is evaluated on the same size of testing data of $\{\beta_i\}_{i=1}^{20}$.}
\small
\centering
\begin{tabular}{c|cccccc}
\hline
         & \multicolumn{2}{c}{$l = 80$ km}  & \multicolumn{2}{c}{$l = 40$ km} & \multicolumn{2}{c}{$l = 20$ km} \\
         \hline
         Training dataset & SSA & MOLHO   & SSA & MOLHO  & SSA & MOLHO \\
         \hline
            $\{\beta_i\}_{i=21}^{200}$ & $7.37 \times 10^{-5}$ & $7.31 \times 10^{-5}$ & $2.09 \times 10^{-4}$ & $1.77 \times 10^{-4}$ & $2.92 \times 10^{-4}$ & $3.04 \times 10^{-4}$ \\
        \hline
            $\{\beta_i\}_{i=21}^{300}$ & $4.84 \times 10^{-5}$ & $4.72 \times 10^{-5}$ & $1.32 \times 10^{-4}$ & $1.47 \times 10^{-4}$ & $2.54 \times 10^{-4}$ & $2.11 \times 10^{-4}$ \\
        \hline
        $\{\beta_i\}_{i=21}^{400}$ & $3.62 \times 10^{-5}$ & $4.09 \times 10^{-5}$ & $1.05 \times 10^{-4}$ & $0.97 \times 10^{-4}$ & $2.28 \times 10^{-4}$ & $1.97 \times 10^{-4}$ \\
\hline
\end{tabular}
\label{table:mismip_training_dataset}
\end{table}

Table \ref{table:mismip_training_dataset} shows that DeepONet converges well with respect to the size $N_{\beta}$ of the training dataset and that using more training data enhances generalization capacity.
The table also shows the impact of the correlation length magnitude on the approximation accuracy. As expected, in order to maintain the same level of accuracy, larger training datasets are required for smaller correlation lengths. Another important piece of information from the table is that DeepONets can approximate with a similar accuracy both the lower-fidelity SSA model and higher-fidelity MOLHO model. 


Taking the case with $\{\beta_i\}_{i=21}^{300}$ as an example, the curves of training and testing losses are plotted in Fig. \ref{fig:mismip-traning}.
The result shows that the DeepONet models converge stably for all three different correlation lengths, and the prediction accuracy on testing cases reaches a plateau after $50000$ epochs. It is observed that the generalization gap\footnote{The difference between a model's performance on training data and its performance on unseen testing data drawn from the same distribution.} remains nearly the same for the data with different correlation lengths when the size of the training dataset is fixed. 

\begin{figure}[ht]
\includegraphics[width=1.0\textwidth]{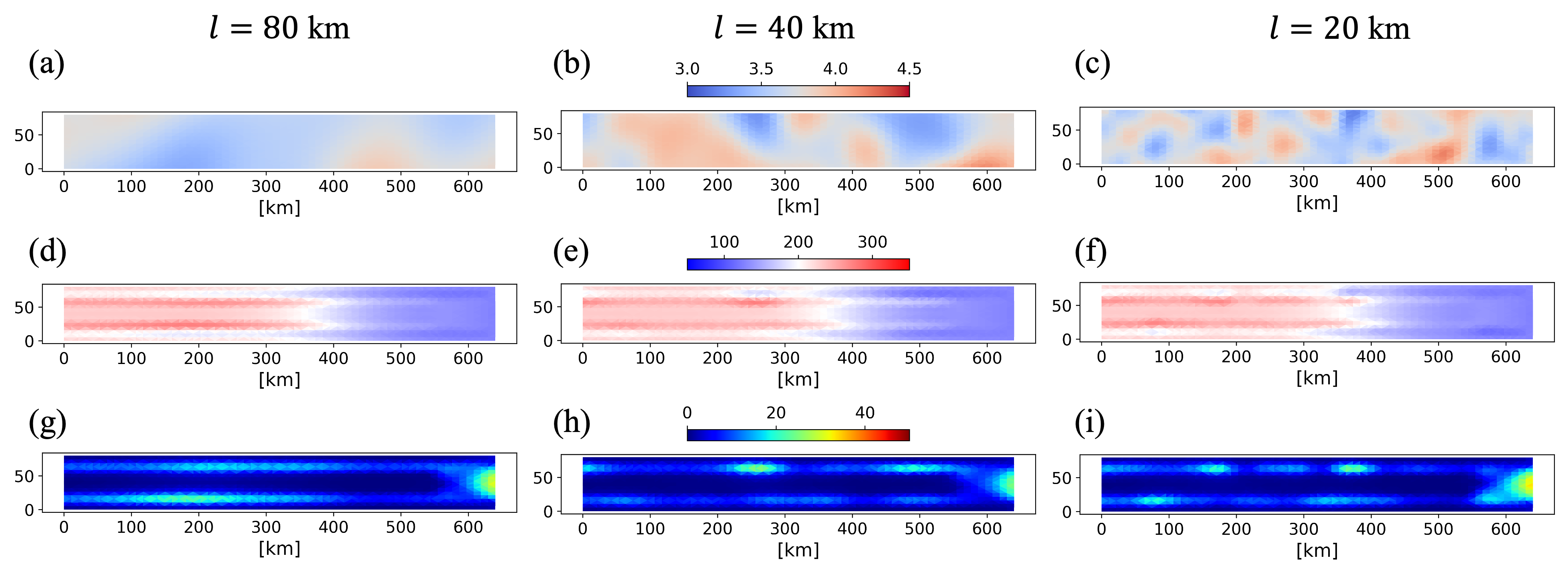}
\caption{The DeepONet prediction at $t = 99$ yr for an exemplary training case ($\beta_{25}$) corresponding to correlation lengths $l = 20, 40, 80$ km. (a) - (c): $\log_{10}(\beta)$; (d) - (f): the thickness $H$; (g)-(i): The modulus of the predicted velocity $|\mathbf{\bar u}^{NN}|$. The relative squared errors are $3.61 \times 10^{-4}$,  $2.57 \times 10^{-3}$, and $7.96 \times 10^{-3}$ for the correlation lengths $l = 80, 40$, and $20$ km, respectively. The simulation data associated with $\{\beta_i\}_{i=21}^{300}$ is used as the training data.}
\label{fig:mismip_beta24}
\end{figure}

The trained DeepONet model $\mathcal{G}_\theta (\beta,H^j)(\mathbf{x})$
is able to predict the velocity field $\mathbf{\bar{u}}^{NN}(\mathbf{x})$ at any time $t^j$ for the given friction field $\beta$ and thickness field $H^j$.
The DeepONet predictions at $t = 99$ yr for an exemplary training case corresponding to correlation lengths $l = 20, 40, 80$ km are presented in Fig. \ref{fig:mismip_beta24}. The results in Fig. \ref{fig:mismip_beta24}(g)-(i) show that more localized features appear in the velocity solution with a smaller correlation length, e.g., the case of $l=20$ km. 
The RSEs between the predicted and reference velocity fields at $t = 99$ yr are $3.61 \times 10^{-4}$,  $2.57 \times 10^{-3}$, and $7.96 \times 10^{-3}$ for the correlation lengths $l = 80, 40$, and $20$ km, respectively, indicating the excellent learning capacity of DeepONet on the training velocity fields.

\begin{figure}[ht]
\centering
\includegraphics[width=0.8\textwidth]{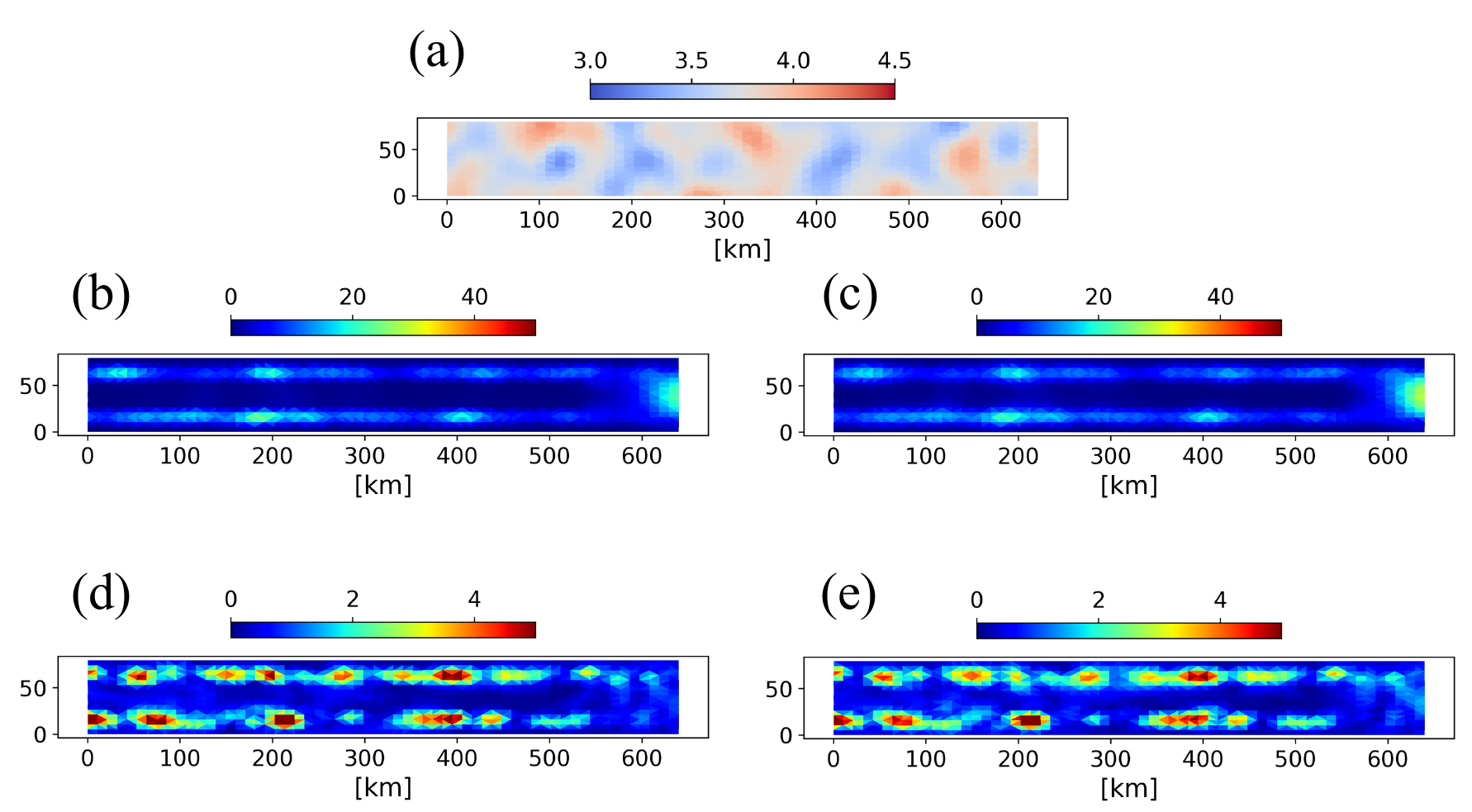}
\caption{The DeepONet prediction for an exemplary test case ($\beta_{6}$) with the correlation length $l = 20$ km: (a) $\log_{10}(\beta)$; (b) and (c) are the maps of reference velocity modulus $|\mathbf{\bar u}|$ at $t = 18$ and $t = 94$ yr, respectively;
(d) and (e) are the point-wise errors of the velocity modulus between the reference and DeepONet predictions $|\mathbf{\bar u} - \mathbf{\bar u}^{NN}|$ at $t = 18$ and $t = 94$ yr, respectively, where the corresponding relative squared errors are $5.67 \times 10^{-2}$ and $4.88 \times 10^{-2}$.}
\label{fig:mismip_l025_beta5}
\end{figure}

To examine the generalization performance, we test the trained DeepONet on an unseen test case ($\beta_6$) with $l = 20$ km at two different time instances, as shown in Fig. \ref{fig:mismip_l025_beta5}. 
The relative squared errors at $t = 18$ and $t = 94$ yr are $5.67 \times 10^{-2}$ and $4.88 \times 10^{-2}$, respectively. We observe that the DeepONet accuracy does not depend significantly on the time $t$ at which the input thickness is evaluated.



Lastly, we investigate the effect of mesh resolution on the DeepONet performance. We use the same $4 \times 300$ DeepONet architecture as before, but we change the size of the input layer to accommodate input data of different resolutions.
Table \ref{table:mismip_training_dataset_mesh} presents the relative squared errors of the DeepONet model against the training dataset $\{\beta_i\}_{i=21}^{300}$ and testing dataset $\{\beta_i\}_{i=1}^{20}$ under different mesh resolutions of $36 \times 9$, $60 \times 15$, and  $100 \times 25$. 
Overall, the accuracy of DeepONet remains comparable for the various mesh resolutions.
The training time for DeepONet under different mesh resolutions is also provided in Table \ref{table:mismip_training_dataset_mesh}, indicating a linear relation between the training time and the size of meshes (i.e., the size of the dataset).

\begin{table}[htbp]
\footnotesize
\caption{MISMIP testcase with the SSA model under different mesh resolutions. The relative squared errors of the DeepONet model against the training dataset $\{\beta_i\}_{i=21}^{300}$ and testing dataset $\{\beta_i\}_{i=1}^{20}$ under various correlation lengths $l$. The clock time used to train DeepONet based on a given mesh resolution remains the same for different correlation lengths.}
\small
\centering
\begin{tabular}{c|c|cccccc}
\hline
          \multicolumn{2}{c}{} &  \multicolumn{2}{c}{$l = 80$ km}  & \multicolumn{2}{c}{$l = 40$ km} & \multicolumn{2}{c}{$l = 20$ km} \\
         \hline
         Mesh resolution & Time & training & testing   & training & testing  & training & testing \\
         \hline
        $36 \times 9$ & $1.13$ hrs & $2.97 \times 10^{-4}$ & $8.02 \times 10^{-3}$ & $0.90 \times 10^{-3}$ & $2.70 \times 10^{-2}$ & $5.28 \times 10^{-3}$ & $6.19 \times 10^{-2}$ \\
        \hline
        $60 \times 15$ & $2.80$ hrs & $3.03 \times 10^{-4}$ & $5.70 \times 10^{-3}$ & $1.14 \times 10^{-3}$ & $1.99 \times 10^{-2}$ & $5.48 \times 10^{-3}$ & $4.25 \times 10^{-2}$ \\
        \hline
        $100 \times 25 $ & $7.24$ hrs & $4.55 \times 10^{-4}$ & $4.02 \times 10^{-3}$ & $1.56 \times 10^{-3}$ & $2.82 \times 10^{-2}$ & $5.44 \times 10^{-3}$ & $4.60 \times 10^{-2}$ \\
\hline
\end{tabular}
\label{table:mismip_training_dataset_mesh}
\end{table}

\section{Hybrid Modeling of Humboldt Glacier} \label{sc:Humboldt}

\begin{figure}[ht]
\centering
\includegraphics[width=.3\textwidth]{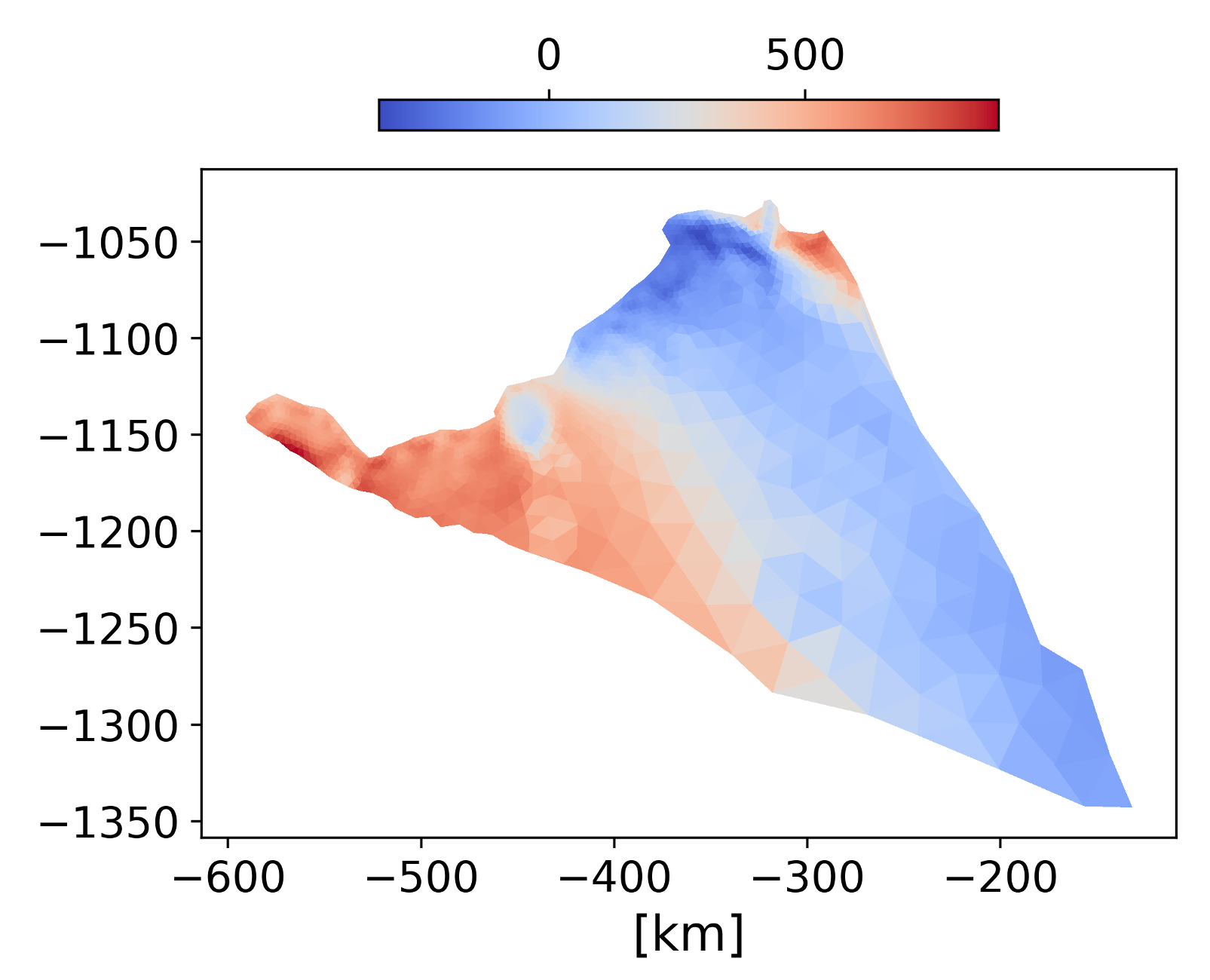}
\includegraphics[width=.3\textwidth]{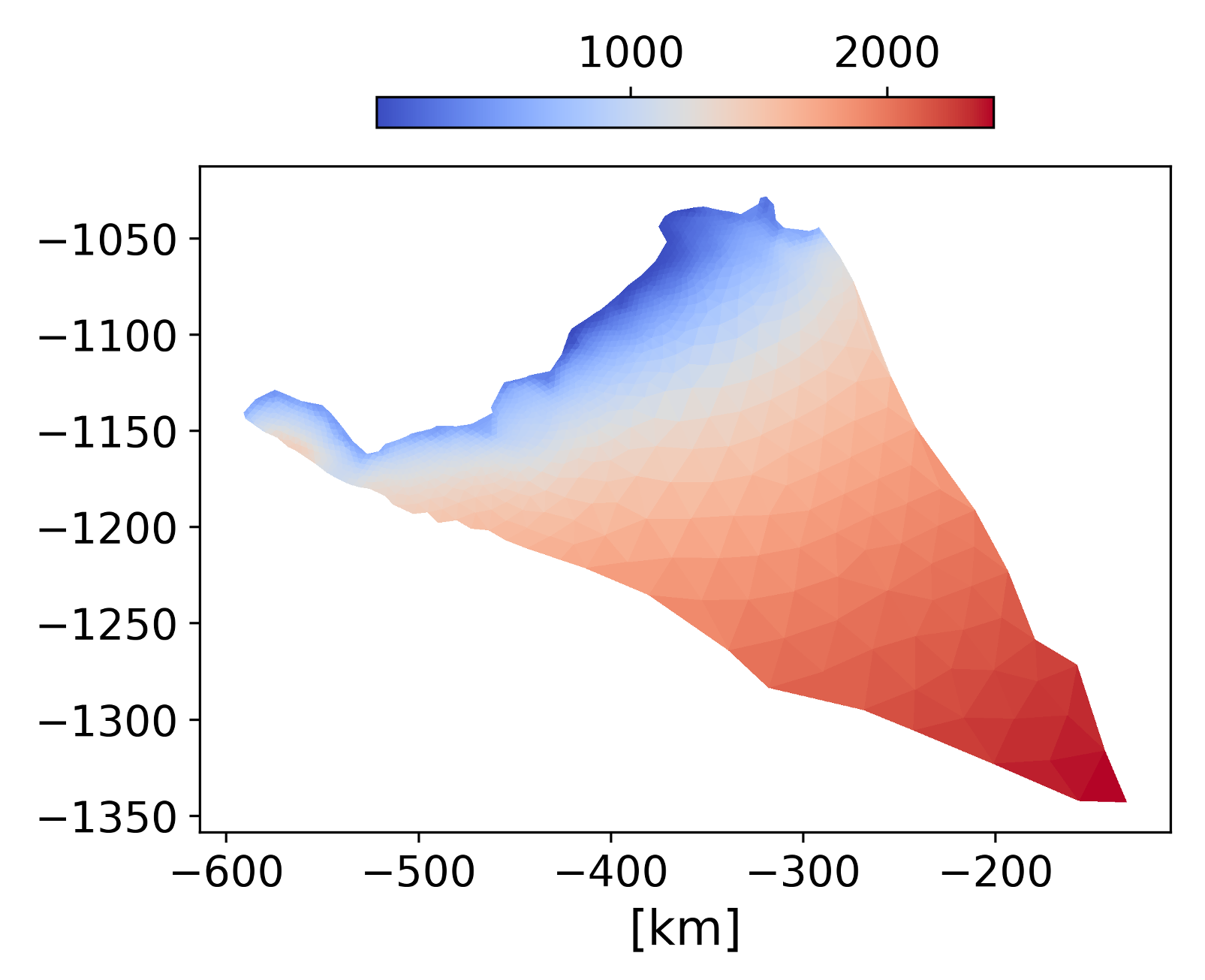}
\includegraphics[width=.3\textwidth]{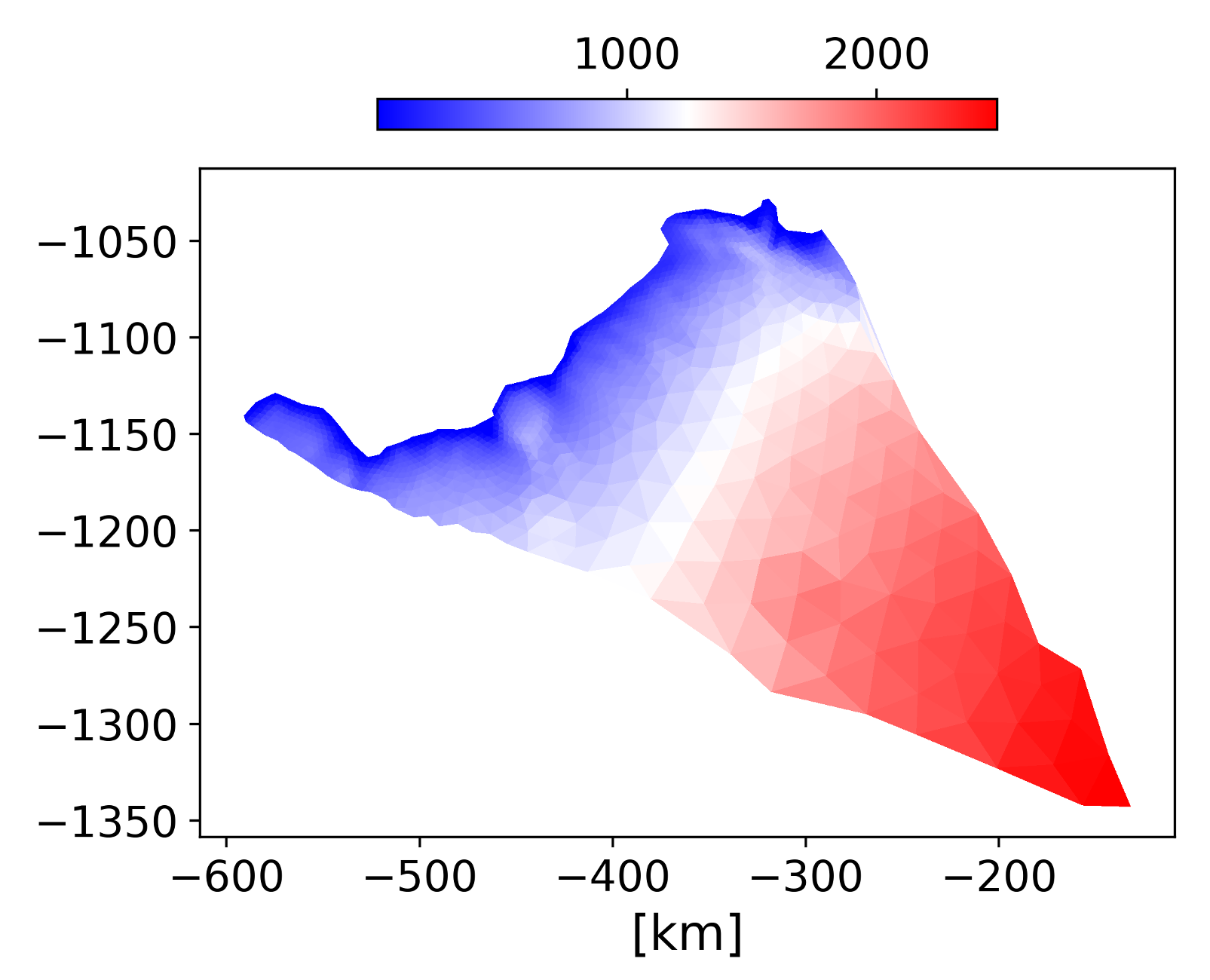}

\caption{Observations for Humboldt glacier for the initial year 2007. Left: bed topography [m] from \cite{morlighem2017}, Center: ice surface elevation [m], 
Right: thickness [m].
Additional details on the collection and processing of these data can be found in \cite{Hillebrand2022}.}
\label{fig:humboldt-glacier}
\end{figure}

\begin{figure}[ht]
\centering
\includegraphics[width=.3\textwidth]{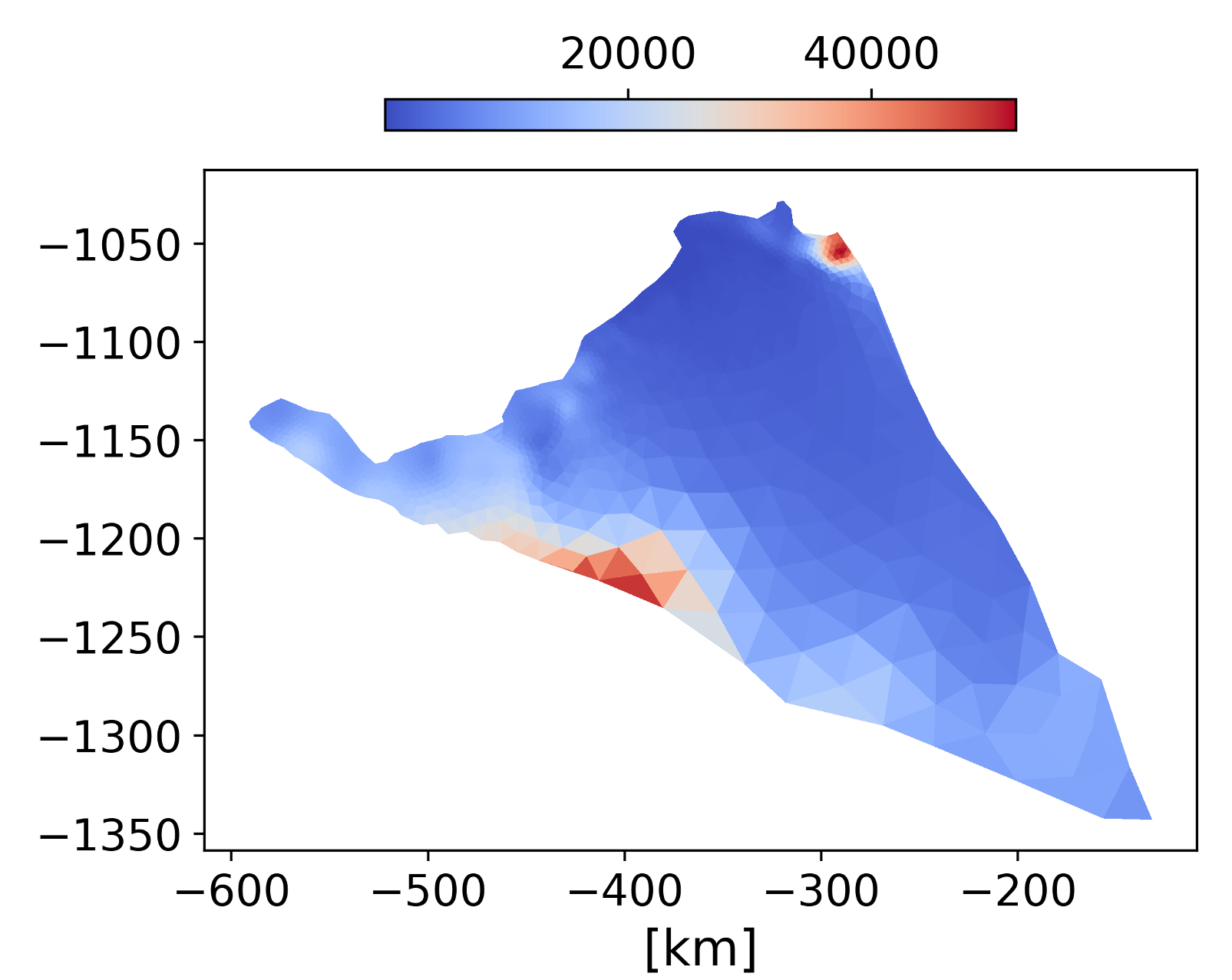}
\includegraphics[width=.3\textwidth]{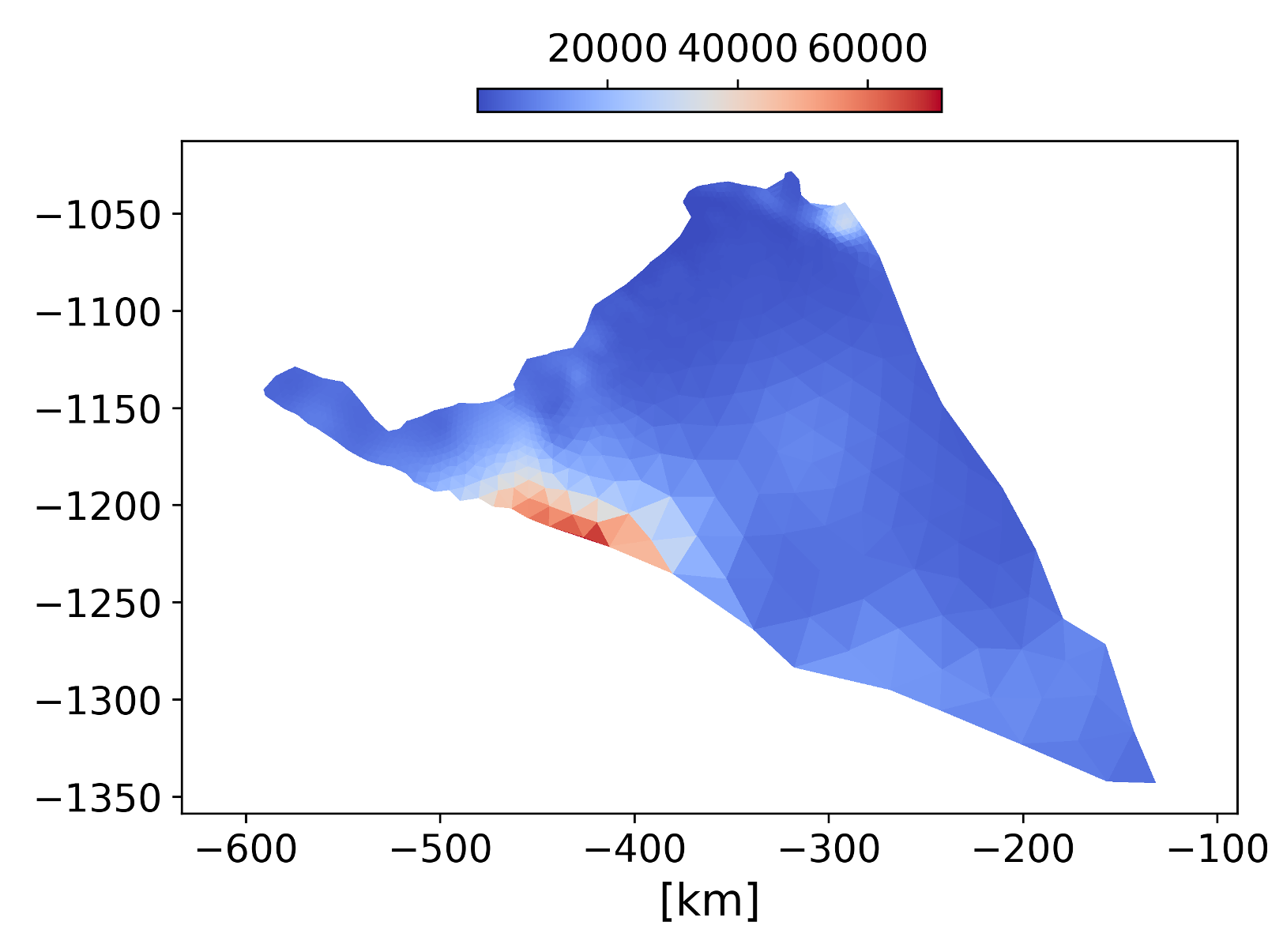}
\includegraphics[width=.3\textwidth]{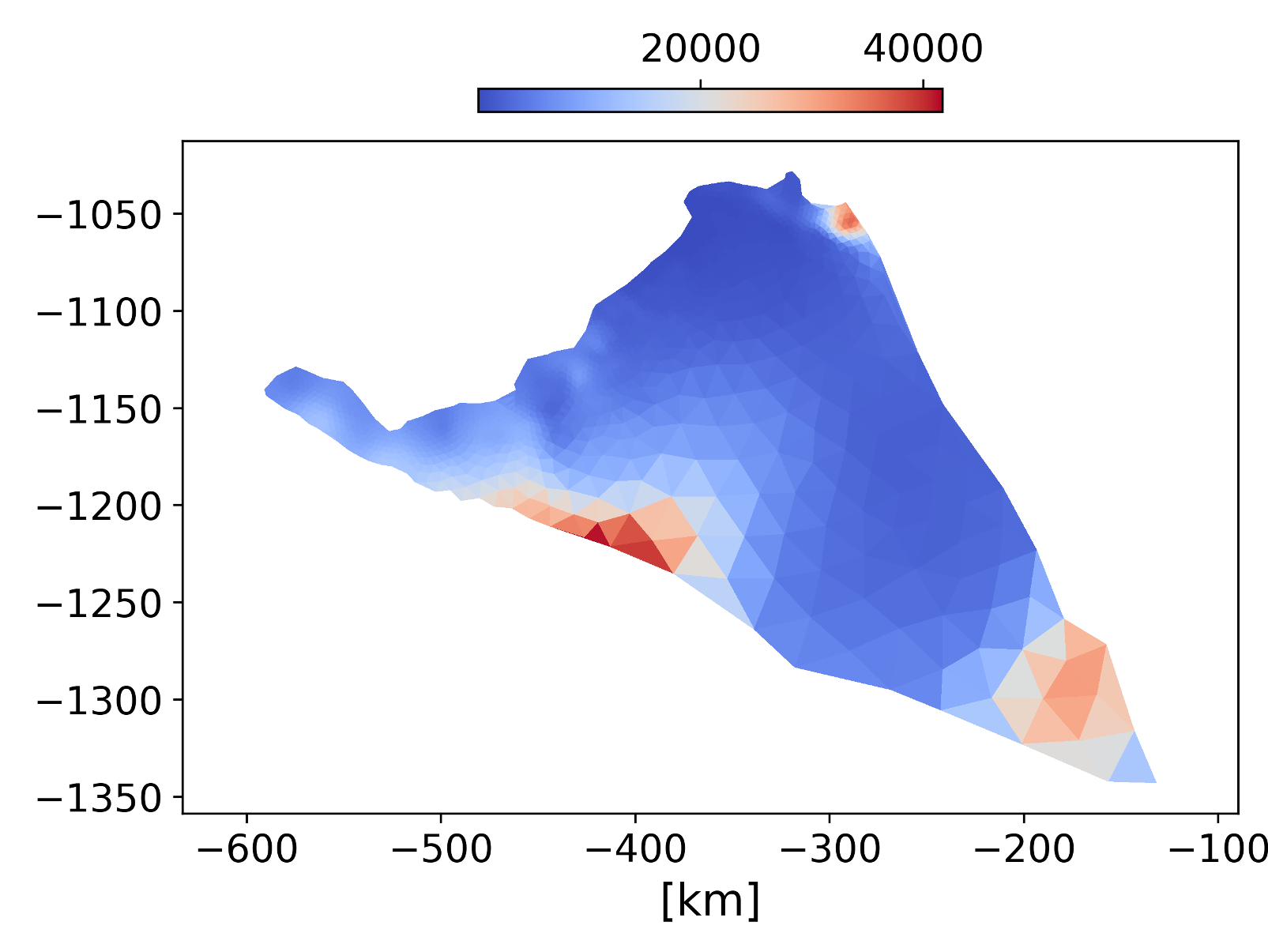}
\caption{Mean value of the basal friction $\bar {\beta}$ (left) and two samples of the basal friction using \eqref{distribution}.Units: [Pa yr / m].} \label{fig:humboldt-beta-samples}
\end{figure}

In this section we consider the Humboldt glacier, one of the largest glaciers in Greenland. In Fig. \ref{fig:humboldt-glacier}, we report the Humboldt bed topography, ice surface elevation and ice thickness obtained from observations, refer to \cite{Hillebrand2022} for details on how these fields are collected and processed. These fields will be use to determine the problem geometry and the initial ice thickness $H^0$.
The mean value $\bar {\beta}$ of the basal friction in \eqref{distribution} is obtained with a PDE-constrained optimization approach \cite{perego2014} where the mismatch between the computed and observed surface velocities are minimized. Fig. \ref{fig:humboldt-beta-samples} shows $\bar {\beta}$ together with a couple of samples of the basal friction from \eqref{distribution}.

Similarly to  the MISMIP case, for each sample of $\beta$, obtained from \eqref{distribution} with correlation length $l=50$~km and scaling $a = 0.2$, the ice finite element flow model is run forward in time for $100$~yr, using a climate forcing generated according to the \emph{Representative Concentration Pathway 2.6} (see \cite{Hillebrand2022} for the problem definition and the data used including the mean basal friction $\bar {\beta}$). The collected thickness and velocity simulation data are used to train the DeepONet model.

\subsection{DeepONet Training}

\begin{figure}[ht]
\centering
\includegraphics[width=1\textwidth]{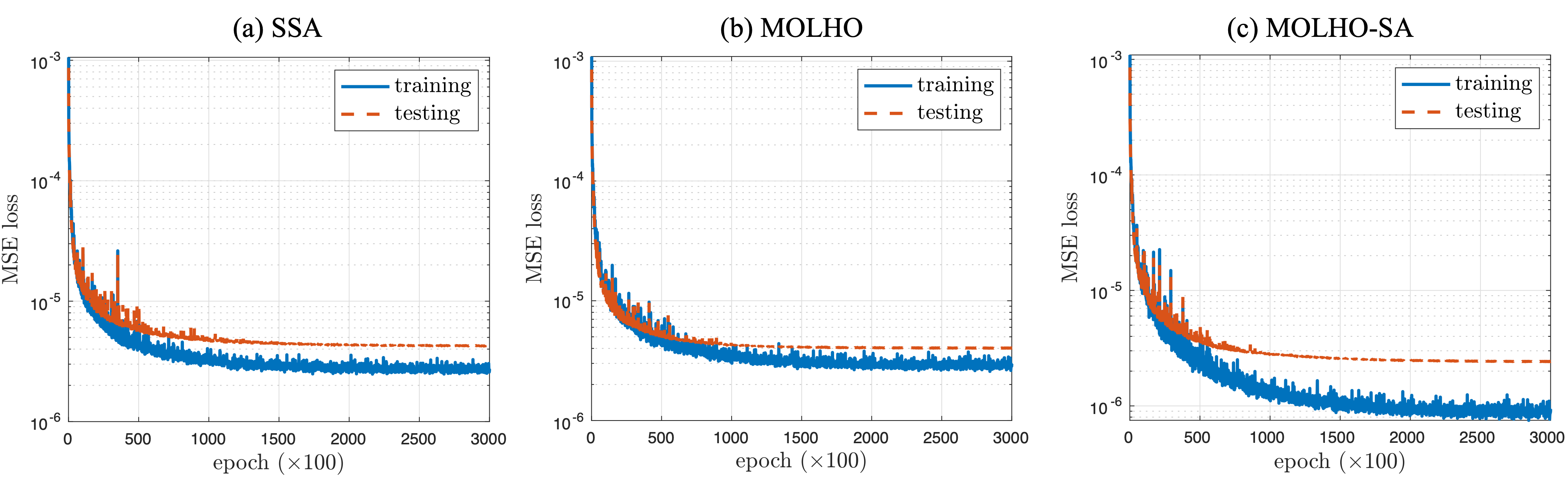}
\caption{The loss plots of DeepONet training for different ice models: (a) SSA; (b) MOLHO; (c) MOLHO with self-adaptive (SA) weighting scheme ($m(\lambda) = \lambda^4$). The simulation data associated with $\{\beta_i\}_{i=21}^{300}$ is used as the training data while $\{\beta_i\}_{i=1}^{20}$ is used as testing data.
At the final epoch ($300,000$), the corresponding testing MSEs of these three DeepONet models are $4.23 \times 10^{-6}$, $4.02 \times 10^{-6}$, and $2.42 \times 10^{-6}$, indicating the enhanced generalization by using the self-adaptive weighting scheme.
}
\label{fig:humboldt-traning_SSA_FO}
\end{figure}

We first evaluate the performance of DeepONet for different ice approximation models (MOLHO and SSA).
Figs. \ref{fig:humboldt-traning_SSA_FO}a-c present the plots of training and testing errors corresponding to three different DeepONet cases, i.e., training with 1) simulation data obtained from the SSA ice model, 2) simulation data obtained from the MOLHO ice model, and 3) simulation data obtained from the MOLHO ice model together with the self-adaptive scheme described in~\eqref{eq:onet_loss_SA}-\eqref{eq:SA_update}. 
At the last epoch ($300,000$), the relative squared errors of these three DeepONet models on the testing data are $3.74 \times 10^{-3}$, $3.59 \times 10^{-3}$, and $2.16 \times 10^{-3}$, respectively. The comparison of results in Figs. \ref{fig:humboldt-traning_SSA_FO}a and b shows that training DeepONet with MOLHO simulation data yields higher prediction accuracy than the low-order SSA data, which is consistent with our observation for the MISMIP testcase.
In the following, we will only consider the MOLHO model, given that it better describes the ice sheet dynamics compared to the SSA model, and it can be well approximated by our DeepONet model. 
We also observe in Fig. \ref{fig:humboldt-traning_SSA_FO}c that the employment of the self-adaptive weighting scheme significantly improves the training and testing performance in the Humboldt glacier testcase, reducing the testing error by $40 \%$.

\begin{figure}[ht]
\centering
\includegraphics[width=1\textwidth]{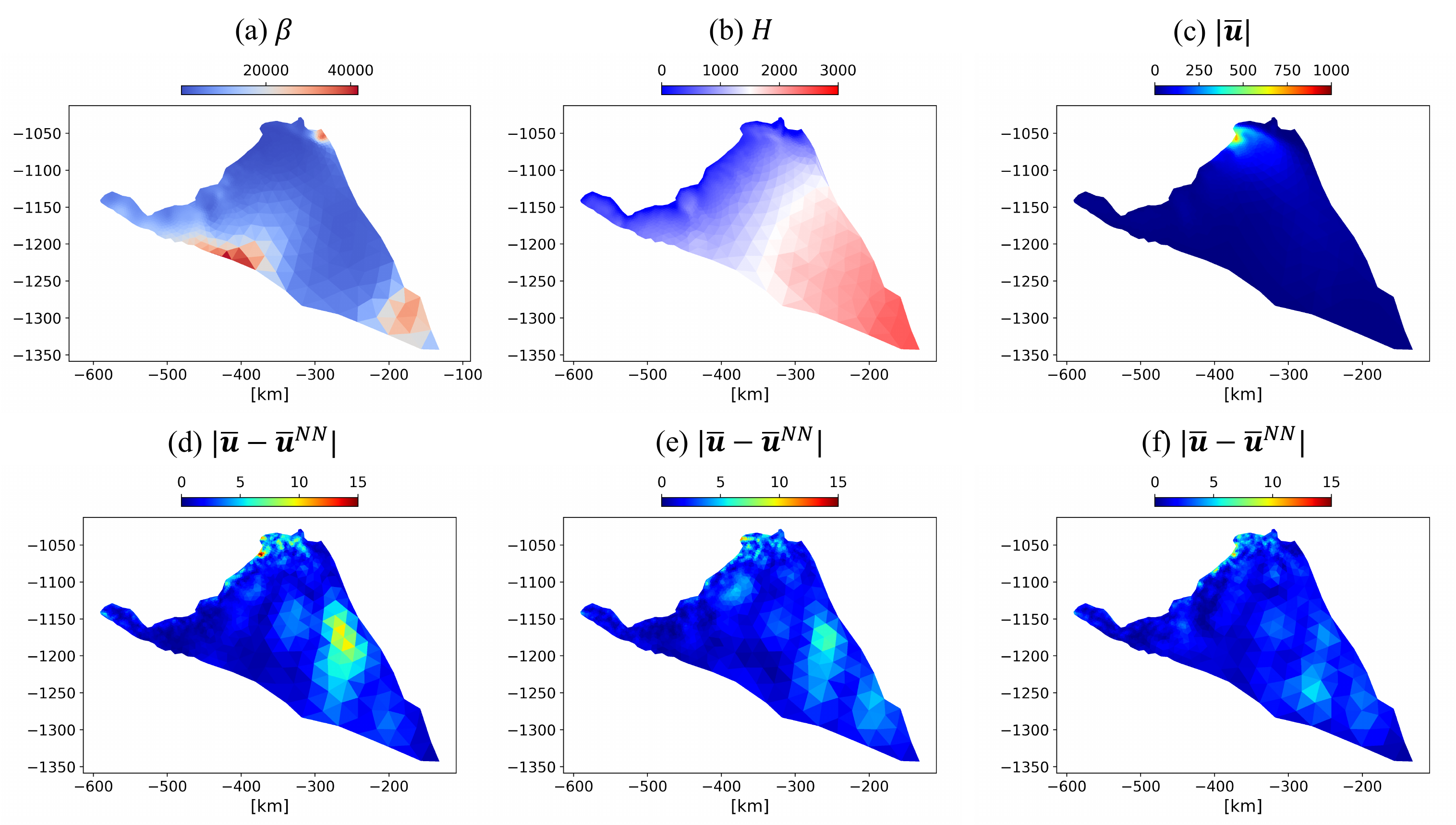}
\caption{The DeepONet prediction for an exemplary training case ($\beta_{23}$) at $t = 99$ yr: (a) basal friction $\beta$ in [Pa yr/m]; (b) Thickness $H$ in [m]; (c) the reference velocity modulus $|\mathbf{\bar u}|$ in [m/yr]; (d) the point-wise errors ([m/yr]) of the DeepONet; (e) the point-wise errors ([m/yr]) of the DeepONet trained with self-adaptive weighting scheme $m(\lambda) = \lambda^2$;
(f) the point-wise errors ([m/yr]) of the DeepONet trained with self-adaptive weighting scheme $m(\lambda) = \lambda^4$. The relative squared errors corresponding to (d)-(f) are $6.29 \times 10^{-4}$, $5.00 \times 10^{-4}$, and $4.18 \times 10^{-4}$, respectively.}
\label{fig:hb_onet_prediction_beta23}
\end{figure}

\begin{figure}[ht]
\centering
\includegraphics[width=1\textwidth]{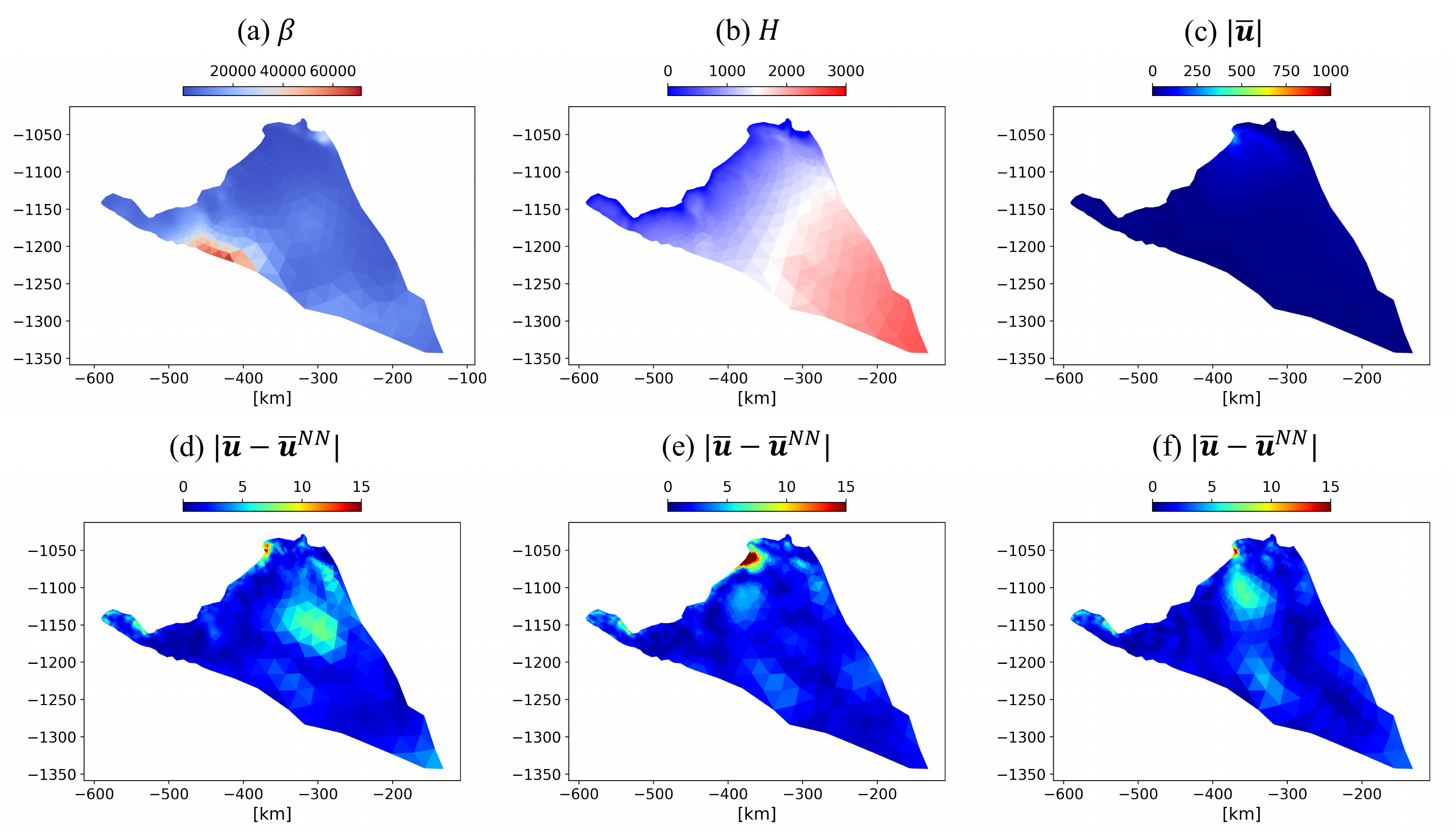}
\caption{The DeepONet prediction for an exemplary testing case ($\beta_{6}$) at $t = 92$ yr: (a) $\beta$ in [Pa yr/m]; (b) Thickness $H$ in [m]; (c) the reference velocity modulus $|\mathbf{\bar u}|$ in [m/yr]; (d) the point-wise errors ([m/yr]) of the DeepONet; (e) the point-wise errors ([m/yr]) of the DeepONet trained with self-adaptive weighting scheme $m(\lambda) = \lambda^2$;
(f) the point-wise errors ([m/yr]) of the DeepONet trained with self-adaptive weighting scheme $m(\lambda) = \lambda^4$. The relative squared errors corresponding to (d)-(f) are $3.36 \times 10^{-3}$, $7.66 \times 10^{-3}$, and $2.88 \times 10^{-3}$, respectively.}
\label{fig:hb_onet_prediction_beta6}
\end{figure}

We further study the impact of using self-adaptive weights in Figs. \ref{fig:hb_onet_prediction_beta23} and \ref{fig:hb_onet_prediction_beta6}, where we show the prediction errors at different $\beta$ samples for different choices of adaptive weighting schemes. In Fig. \ref{fig:hb_onet_prediction_beta23} we report the results for a $\beta$ sample taken from the training dataset, whereas in Fig. \ref{fig:hb_onet_prediction_beta6} we consider a sample from the testing dataset. In both cases, the DeepONet model trained with the self-adaptive weighting scheme with $m(\lambda)=\lambda^4$ yields the best performance, which is consistent with the results in Fig. \ref{fig:humboldt-traning_SSA_FO}. The self-adaptive weighting scheme especially helps mitigate the prediction errors in the interior of the domain and the region at the outlet (i.e., northwest) region.
Given the improved prediction, in the following sections we will present the DeepONet models trained with the self-adaptive weighting scheme with $m(\lambda)=\lambda^4.$


\subsection{Hybrid model: DeepONet embedded in finite element solver}
In this section we study the accuracy and cost of the hybrid ice flow model with respect to the finite element model. As explained in Section \ref{sc:computational_models}, the hybrid model approximates at each time step the operator $\mathcal{G}$ with the trained DeepONet model $\mathcal{G}_\theta$. Because the DeepONet approximation is much cheaper than the finite element approximation, the hybrid solver is significantly more efficient than a traditional finite element solver. We study the approximation properties and computational savings of using the hybrid model for computing the evolution of the Humboldt glacier thickness over time, and then focus in particular on how well the hybrid model can approximate the glacier mass change. We finally show how the hybrid model can be used to produce statistics of the glacier mass loss.

\subsubsection{Thickness evolution over time}
\begin{figure}[ht]
\centering
\includegraphics[width=.8\textwidth]{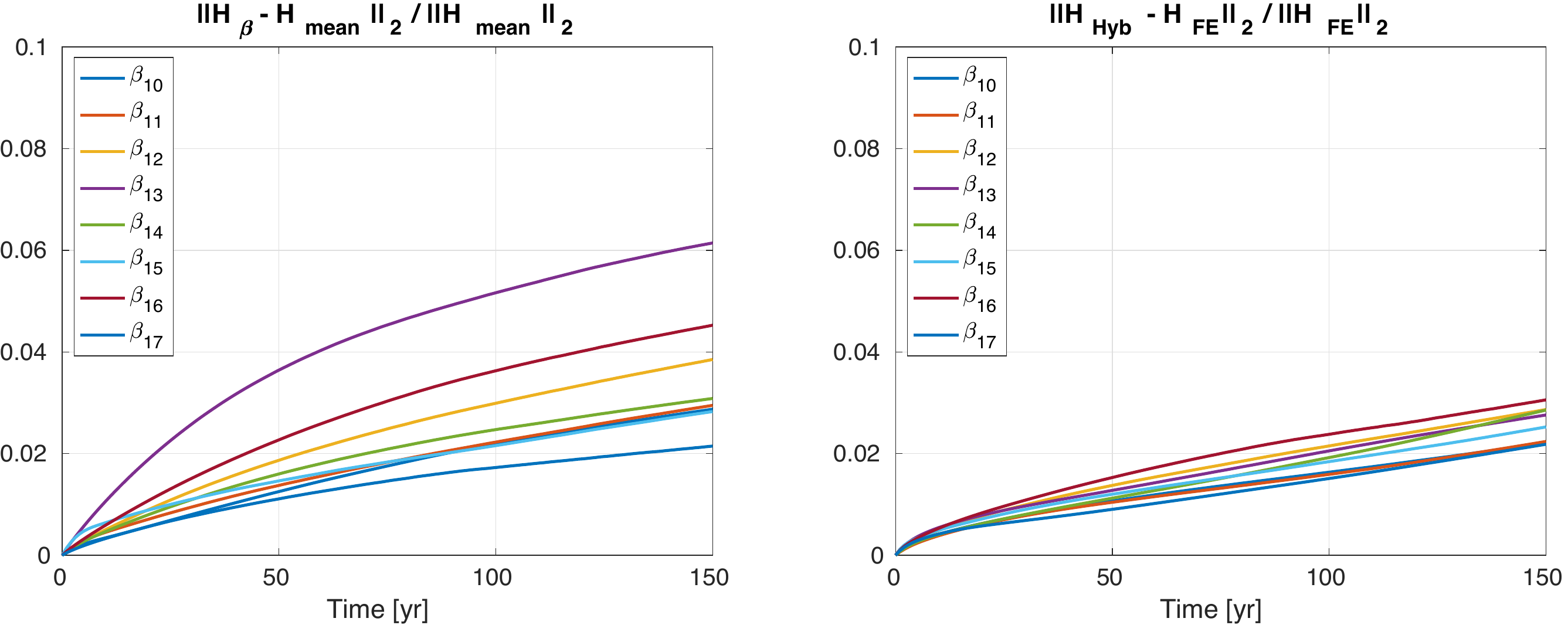}
\caption{Left: relative difference over time between the ice thickness $H_{\beta}$ associated to sample $\beta$ and the mean ice thickness. Right: relative difference over time between the ice thickness computed with the finite element model and the hybrid model.} \label{fig:humboldt-thickness}. 
\end{figure}

In this section we compare the ice thickness computed with the finite-element model, and with the hybrid model. We take $8$ samples of beta (not used to train the DeepONet) from distribution \eqref{distribution}. We then run the finite-element and the hybrid models for $150$ years. Results of the comparison are shown in Fig. \ref{fig:humboldt-thickness}. The plot on the left shows the variability of the thickness, over time, with respect to the samples of $\beta$, using the same model. The plot on the right shows the relative difference between the ice thickness computed with the finite-element model and the one computed with the hybrid model. The relative differences due to the models are significantly smaller than the variability with respect to the different samples. Moreover, for $t < 100$ years, which is the period used for training the DeepONet, the relative differences between the two models are small, $3$\% at most. Differences increase in the extrapolation region ($100 - 150$ years), however the increase is mostly linear, which signifies robustness of the hybrid approximation.

\subsubsection{Glacier mass-loss over time}
\begin{figure}[ht]
\centering
\includegraphics[width=.8\textwidth]{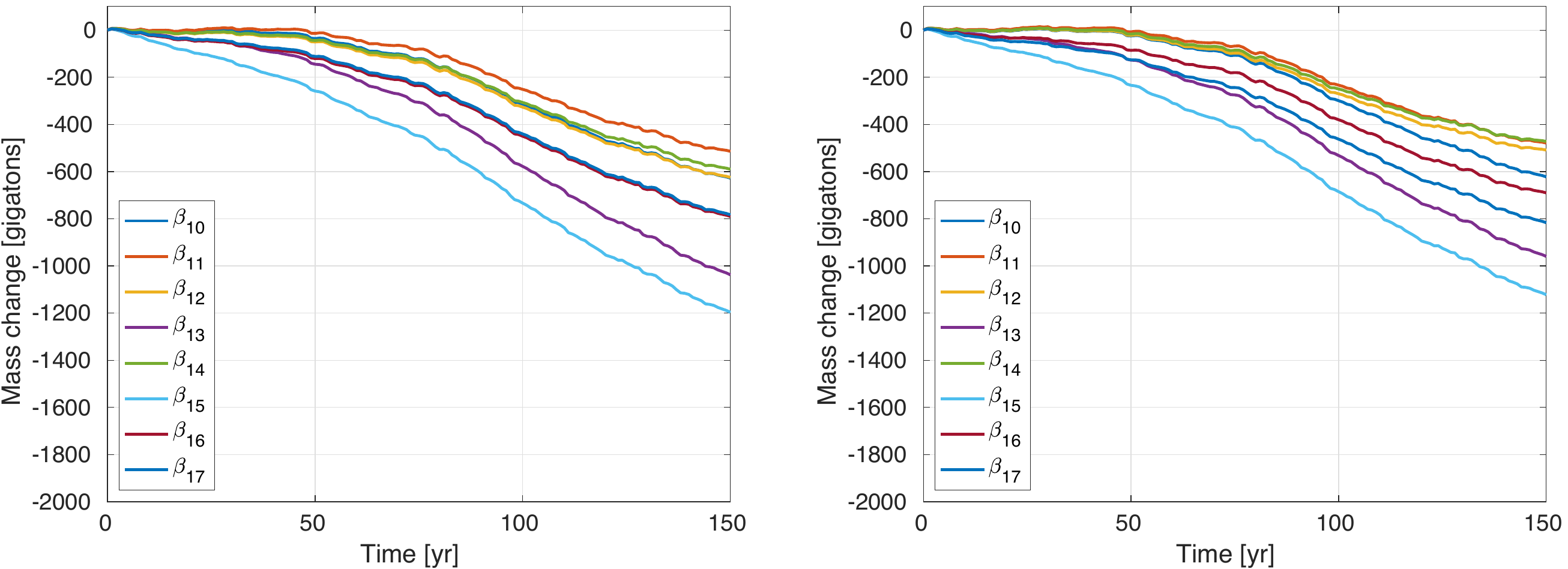}
\caption{Mass change [gigatons] over time for different samples of the basal friction coefficient computed using the finite element model (left) and the hybrid model (right).} \label{fig:humboldt-mass-loss}
\end{figure}

As explained in the introduction, the mass change of a glacier over the years is one of the most important quantities of interest in ice sheet modeling because it directly affects the net amount of water added to the oceans and hence the potential sea level rise. In this work, we compute the mass of the glacier only considering the ice that is above flotation, because changes in the mass of ice that is afloat do not affect the sea level; for details, see \cite{goelzer_2020}. In Fig. \ref{fig:humboldt-mass-loss}, we show the mass change (mass at time $t$ minus mass at time $t_0 = 0$) as a function of time for the same samples of the basal friction used for Fig. \ref{fig:humboldt-thickness}. While there are some small discrepancies between the finite-element and hybrid models, the two model are in very good agreement overall, especially in the first $100$ years, which are within the period of ice simulation data used for training the DeepONet model, with the largest difference being $\approx 10$\%. We also note that the qualitative behaviors of the two models are very similar  in the extrapolation region ($100-150$ years).

\subsubsection{Computing statistics on quantity of interest using Hybrid model}
\begin{figure}[ht]
\centering
\includegraphics[width=0.8\textwidth]{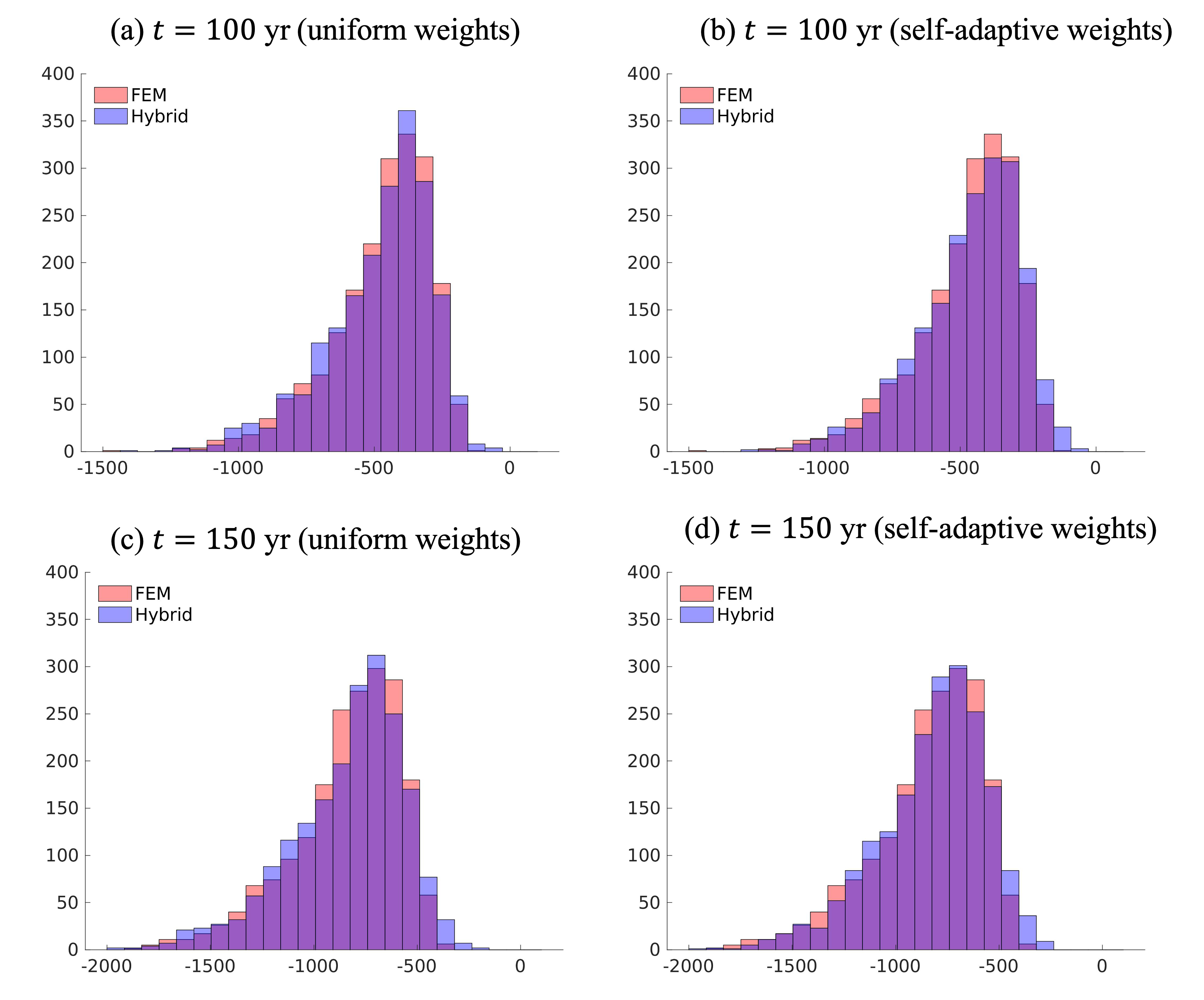}
\caption{Histogram of the distribution of the Humboldt mass change over a period of 100 years and 150 years. The histogram has been generated by simulating the mass change corresponding to $2000$ samples from distribution \eqref{distribution}. The DeepONet used for the results on the right (b and d) has been trained using self-adaptive weights, whereas uniform weights have been used for the results on the left (a and c).} 
\label{histograms}
\end{figure}

Finally, we demonstrate how the hybrid model can be effectively used to compute statistics of the glacier mass change. We take unseen $2000$ samples of $\beta$ from distribution \eqref{distribution}, and run both the hybrid model and the finite-element model for $100$ years and $150$ years for each sample. We then compute the glacier mass change (using only the ice above flotation) and show histograms (Fig. \ref{histograms}) of the mass-change distribution, comparing the differences between the reference finite element model and the hybrid model.  The results demonstrate that the hybrid model can accurately compute the statistics of mass change, and therefore has the potential to be used to significantly accelerate the uncertainty quantification analysis for sea-level projections due to ice-sheet mass change. The discrepancies between the results computed with the reference finite element model and the hybrid model are likely small in practical applications, and, if needed, they can be corrected using a multifidelity approach where the hybrid model is used as low-fidelity model and the finite-element model as the high-fidelity model; see e.g., \cite{Peherstorfer_WG_SIAM_2016}. The figure also shows the impact of training the DeepONets using self-adaptive weights and uniform weights. It seems that the use of self-adaptive weights in training can lead to a small bias in the hybrid modeling to underestimate the mass loss. More investigation is needed to understand the cause of this bias and to confirm that this phenomenon is general and not specific to this particular glacier and the settings we used.

\subsubsection{Computational saving using Hybrid model} \label{sc:savings}
Table \ref{table:computational_time} shows the computational times for running the finite element and hybrid models, when using the MOLHO approximation. Overall, we see almost a 5-fold speedup when using the hybrid model over the finite element model. The total computational costs includes time to allocate memory, initialize data and for intput/output. If we only consider the time to solve the coupled model system \eqref{time-disc-thickness}, we have a 11-fold speedup. The evaluation of the DeepONet takes only 4.99s of the 9.46s taken to solve the hybrid model. We believe that there is margin for improvement in real applications.
While we trained the DeepONet on GPUs, our prototype FEniCS code can only run on CPUs, therefore the results in this section refers to simulations run on CPUs. We expect that the DeepONet would benefit more from running on GPUs than a the classic finite element model, because it is still challenging to efficiently run implicit nonlinear solvers on GPUs (see \cite{watkins2022}, in the context of a production ice sheet model), whereas modern machine learning code can take full advantage of GPUs.
The cost of the finite element solver scales with increasing mesh resolutions whereas DeepONet can maintain the same level of predictive accuracy and efficiency for various mesh resolutions (as shown in Table \ref{table:mismip_training_dataset_mesh}).
Moreover, we expect that an hybrid model would be significantly more efficient, compared to the corresponding finite element code, when higher-order approximations of the velocity solver are considered. In fact, a Stokes solver can be an order of magnitude slower than the MOLHO model considered here, whereas we expect the cost of the DeepONet to be fairly independent from the model chosen for the velocity solver, as we observed when comparing the SSA and the MOLHO DeepONet models.

\begin{table}[htbp]
\footnotesize
\caption{Comparison of computational time per sample between the finite-element and hybrid models when using MOLHO model for the velocity solver. The provided average times are estimated based on 50 simulations with different friction fields.}
\small
\centering
\begin{tabular}{c|ccc}
\hline 
        Times per sample (s) & Total   & Solving Eq. \eqref{time-disc-thickness} \\
        \hline
        Finite-element model  & 123.30   &   105.20 \\
        Hybrid model   &  24.15   &  9.46    \\
        \hline
        Ratio   & 19.59 \%  & 8.99\%     \\
        \hline
\end{tabular}
\label{table:computational_time}
\end{table}

\section{Summary}\label{sc:summary}
We developed a hybrid model for ice sheet dynamics by combining a classic finite-element discretization for the ice thickness equation with a DeepONet approximation of the ice momentum equation, which is the most expensive part of a traditional ice sheet computational model. A distinctive feature of our hybrid model is that it can handle high-dimensional parameter spaces, which is critical for accounting for the uncertainty in parameter fields like the basal friction coefficient. We demonstrated that the hybrid model can accurately compute the dynamics of a real glacier an order of magnitude faster than a traditional ice sheet model. As explained in Section \ref{sc:savings}, the computational savings are likely to be larger when using production ice-sheet codes. Moreover, we showed that the hybrid model produces accurate statistics of the mass loss of the Humboldt glacier over a period of one hundred years and can therefore be used to accelerate uncertainty quantification analysis of sea-level projections due to ice sheets. Future research directions include scaling up our approach to target larger problems, such as using higher-resolution data or targeting the evolution of the entire Greenland ice sheet, and performing uncertainty quantification analysis using the hybrid model.

\section{Acknowledgements}
The authors wish to thank L. Lu for helpful discussions, K. C. Sockwell for co-developing the ice-sheet code, and T. Hillebrand for generating the Humboldt grid. 

The work is supported by the U.S. Department of Energy, Advanced Scientific Computing Research program, under the Physics-Informed Learning Machines for Multiscale and Multiphysics Problems (PhILMs) project and under the SciDAC-BER Probabilistic Sea Level Projections from Ice-Sheets and Earth System Models (ProSPect) partnership.
The authors also acknowledge the support from the UMII Seed Grant and the Minnesota Supercomputing Institute (MSI) at the University of Minnesota for providing resources that contributed to the research results reported within this paper.

Sandia National Laboratories is a multimission laboratory managed and operated by National Technology and Engineering Solutions of Sandia, LLC., a wholly owned subsidiary of Honeywell International, Inc., for the U.S. Department of Energy’s National Nuclear Security Administration under contract DE-NA-0003525.

Pacific Northwest National Laboratory (PNNL) is a multi-program national laboratory operated for the U.S. Department of Energy (DOE) by Battelle Memorial Institute under Contract No. DE-AC05-76RL01830. The computational work was performed with resources from PNNL Institutional Computing at Pacific Northwest National Laboratory. 

This paper describes objective technical results and analysis. Any subjective views or opinions that might be expressed in the paper do not necessarily represent the views of the  U.S. Department of Energy or the United States Government. 
\clearpage

\bibliographystyle{elsarticle-num} 
\bibliography{refs}


\end{document}